\UseRawInputEncoding
\documentclass[%
 reprint,
 superscriptaddress,
 amsmath,amssymb,
 aip,
]{revtex4-2}

\usepackage{graphicx}
\usepackage{dcolumn}
\usepackage{bm}
\usepackage[colorlinks=true,urlcolor=blue,breaklinks=true]{hyperref}
\usepackage{newtxtext,newtxmath}
\usepackage[dvipsnames]{xcolor}
\usepackage{braket}
\usepackage{booktabs}
\usepackage{placeins}

\begin{document}

\preprint{APS/123-QED}

\title{Modeling of flopping-mode spin qubits: beyond the two-site model}

\author{Ashutosh Kinikar}\email{ashutosh.kinikar@imec.be}
\affiliation{Imec, Kapeldreef 75, 3001 Heverlee, Belgium}
\affiliation{Instituut voor Theoretische Fysica, KU Leuven, Celestijnenlaan 200D, 3001 Leuven, Belgium}

\author{Vukan Levajac}
\affiliation{Imec, Kapeldreef 75, 3001 Heverlee, Belgium}
\affiliation{Instituut voor Theoretische Fysica, KU Leuven, Celestijnenlaan 200D, 3001 Leuven, Belgium}

\author{Kristof Moors}
\affiliation{Imec, Kapeldreef 75, 3001 Heverlee, Belgium}

\author{George Simion}
\affiliation{Imec, Kapeldreef 75, 3001 Heverlee, Belgium}

\author{M\'onica Benito}
\affiliation{Institute of Physics, University of Augsburg, Augsburg, 86159, Germany}
\affiliation{Center for Advanced Analytics and Predictive Sciences, University of Augsburg, 86135 Augsburg, Germany}

\author{Bart~Sor\'ee}\email{bart.soree@imec.be}
\affiliation{Imec, Kapeldreef 75, 3001 Heverlee, Belgium}
\affiliation{Department of Electrical Engineering, KU Leuven, Kasteelpark Arenberg 10, 3001 Heverlee, Belgium}
\affiliation{Department of Physics, University of Antwerp, Groenenborgerlaan 171, 2020 Antwerp, Belgium}

\date{\today}

\begin{abstract}
    We present a flexible modeling framework for flopping-mode spin qubits that captures the spatial structure of the double-well confinement and magnetic-field-gradient profile going beyond conventional low-energy descriptions.
    By using this approach, we simulate electric dipole spin resonance-based single-qubit control and evaluate the frequency and spectral purity of the Rabi oscillations across different parameter regimes.
    Our analysis reveals a fundamental tradeoff between fast electrical driving and clean single-mode Rabi oscillations, and demonstrates that the standard two-site low-energy approximation can overestimate the Rabi frequency by up to $\sim$ 20\% in certain parameter regimes.
    We also investigate two-qubit control by considering two capacitively coupled flopping-mode qubits and derive the corresponding exchange interaction with an appropriately restricted configuration interaction treatment.
    Our approach reveals the interplay between the spatial profile of the double-well confinement, magnetic field gradient, and Coulomb interaction, which together govern the effective exchange coupling strength.
   Our spatially resolved modelling framework enables efficient exploration of double-well confinement parameters and magnetic field gradient profiles, enabling a transparent mapping from spatial device properties to flopping-mode qubit parameters and quality metrics.
\end{abstract}

\maketitle

\section{Introduction}
\label{sec:introduction}
The spin of an electron or hole in a gate-defined quantum dot in the presence of a constant magnetic field forms a natural two-level system with a long coherence time.~\cite{Loss1998, Burkard2023}
This qubit system is ideally suited for large-scale integration with CMOS foundry-compatible fabrication processes on silicon or germanium.~\cite{Zajac2016, Veldhorst2017, Kuenne2024, Li2026}
Coherent control between the two spin states can be achieved by applying an oscillating magnetic field perpendicular to the constant magnetic field, a technique known as electron spin resonance (ESR).~\cite{Koppens2006, Pla2012, Veldhorst2014}
ESR is conceptually simple, but requires high-frequency magnetic fields that act globally and therefore complicate single-qubit addressability~\cite{Vandersypen2017}. Moreover, such high-frequency fields can cause significant power dissipation and heating~\cite{Takeda2016}. In general, these factors limit the scalability of the ESR control technique.

To overcome these challenges, electric dipole spin resonance (EDSR) is an interesting alternative. EDSR enables electric spin control of an individual qubit by coupling its spin and orbital degrees of freedom.~\cite{RashbaEfros2003}
In the presence of a transverse magnetic field gradient, an oscillating electric field displaces the electron, which effectively creates an oscillating transverse magnetic field that acts on its spin, as with ESR.~\cite{PioroLadriere2008, Unseld2025}
A particularly efficient realization of the EDSR principle is the \emph{flopping-mode} (FM) spin qubit (see Fig.~\ref{fig:FM_qubits}(a)).~\cite{Benito2019, Croot2020, Teske2023, Hu2023, Young2025}
In a FM qubit, an electron is confined in a double-well potential and its charge and spin are delocalized over the two wells.
This delocalization increases the effective electric dipole moment of the electron.
This makes FM qubits attractive for fast single-qubit control, as it enhances the response of the spin to an EDSR drive in the presence of a transverse magnetic field gradient. Also, the large electric dipoles of FM qubits allow for long-range coupling mechanisms, for example via spin-photon interactions in cavity architectures.~\cite{Borjans2019, HarveyCollard2022, Dijkema2024, Jiang2025}

A common approach for quantum dot spin qubit modeling is to consider a low-energy description on a lattice, with effective sites for the quantum dot confinement and phenomenological tunnel coupling, detuning and exchange parameters. 
Such low-energy lattice descriptions have been used to study a range of quantum dot properties, including charge stability diagrams and single-electron transport through coupled dot systems.~\cite{Wang2011, Rassekh2022}
Although such low-energy descriptions of FM qubits are insightful and computationally efficient,~\cite{Benito2019, Cayao2020} they do not offer a direct connection between microscopic device properties and qubit performance, which limits their predictive power for design analysis and optimization.
The performance of FM qubits is intrinsically tied to various microscopic device properties, such as the well separation, inter-well barrier height, confinement strength, and the magnetic field gradient profile.
These properties determine orbital energies, tunnel couplings, electric dipole moments, effective spin-charge mixing, and effective exchange strengths, which in turn govern the single-qubit and two-qubit control performance. Furthermore, these quantities also influence the susceptibility of the qubit to environmental noise and leakage into states outside of the computational basis. 

In this work, we apply a flexible modeling framework that captures the important spatial properties of FM qubits and still allows for computationally efficient modeling, avoiding
more costly finite-element-method simulations. ~\cite{Mohiyaddin2019, Niquet2020, Beaudoin2022, Shehata2023, Costa2023, Pedicini2025}
Starting from a spatially resolved double-well confinement potential and magnetic field gradient profile, we construct an orbital basis tailored to the underlying geometry.
For this, we employ a central basis set with Hermite polynomials,~\cite{Shehata2023} which enables analytical evaluation of matrix elements and efficient diagonalization.
For the exchange interaction between FM qubits, we consider an appropriately restricted configuration interaction treatment based on this orbital basis.
We are thus equipped to analyze both single- and two-qubit control of FM qubits with our approach.
For single-qubit control, we evaluate the EDSR-driven dynamics of an FM qubit across a wide range of double-well confinement regimes and map the Rabi oscillation frequency and spectral purity of the oscillations. For two-qubit control, we consider two neighboring FM qubits that are capacitively coupled, and we extract the exchange interaction over a range of double-well confinement parameters and separations between the qubits.~\cite{Cayao2020}
For both types of control, our approach reveals the impact of spatially resolved device parameters on qubit performance and uncovers quantitative and qualitative differences with respect to effective two-site descriptions, demonstrating the importance of retaining the full spatial structure of the device in FM qubit modeling.

The paper is organized as follows. In Sec.~\ref{sec:model}, we introduce the spatially resolved modeling framework for the single FM qubit and its extension to two capacitively coupled FM qubits. In Sec.~\ref{sec:Results}, we present the main results: in Sec.~\ref{sec:FM_microscopics} we discuss the FM qubit spectrum and add an in-depth analysis of its spatial properties, in Sec.~\ref{sec:EDSR} we analyze EDSR-based single-qubit control including Rabi frequencies and the spectral purity of the oscillations, and in Sec.~\ref{sec:exchange} we investigate the exchange interaction between two neighboring FM qubits. We discuss some specifics and extensions of our modeling approach in Sec.~\ref{sec:discussion} and conclude in Sec.~\ref{sec:conclusion}. Technical details, complementary low-energy analysis, and supporting analytical and numerical results are provided in the Appendices.

\section{Model}
\label{sec:model}
\subsection{Single flopping-mode qubit}
\label{subsec:single_FM_model}
\begin{figure}[tb]
    \includegraphics[width=\linewidth]{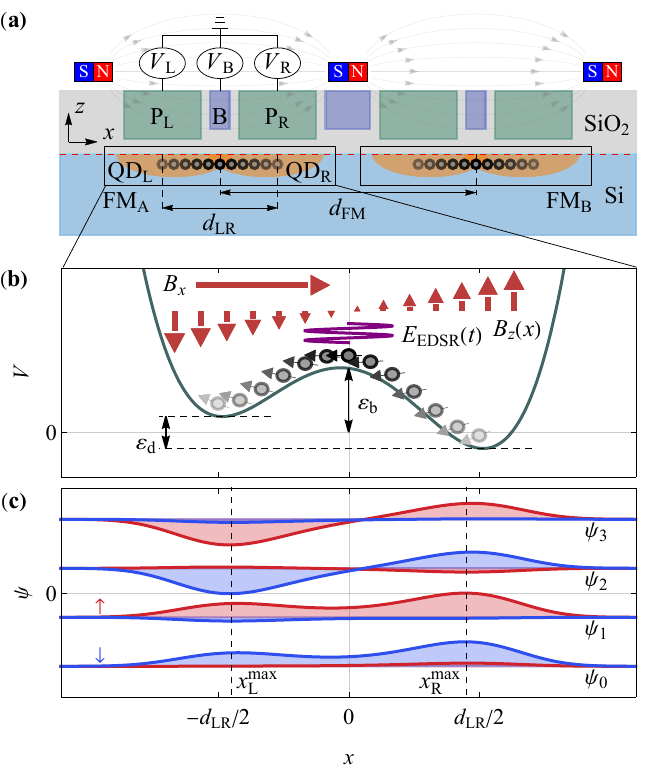}
    \caption{
        \textbf{Flopping-mode qubits}.
        \textbf{(a)} Schematic of two neighboring flopping-mode spin qubits in a Si metal-oxide-semiconductor gate stack, with relevant gate voltages indicated for the left qubit ($\text{FM}_\text{A}$).
        \textbf{(b)} Double-well confinement potential of a flopping-mode qubit, with detuning $\varepsilon_\mathrm{d}$ (controlled by the gate voltages on the plungers $V_\text{L}, V_\text{R}$), barrier height $\varepsilon_\mathrm{b}$ (controlled by the gate voltage on the barrier $V_\text{B}$), a constant magnetic field $B_x$ in the double-well confinement ($x$) direction, and a variable field $B_z(x)$ along the transverse ($z$) direction varying with a linear gradient along the $x$ direction (generated by micromagnets). An oscillating electric field along the longitudinal ($x$) direction enables EDSR, and the barrier gate voltage controls the exchange strength through capacitive coupling.
        \textbf{(c)} The spatially resolved wave functions of the four lowest-energy states (labeled $0\--3$).
    }
    \label{fig:FM_qubits}
\end{figure}

We restrict our description of a single FM qubit to an effective one-dimensional model for an electron with effective mass $m_\mathrm{e}^\ast$ along the confinement axis of the double-well potential (see Fig.~\ref{fig:FM_qubits}). This restriction captures the dominant orbital dynamics relevant for FM operation. Our modeling framework can be extended to include more microscopic details (e.g. transverse confinement, valleys) in a straightforward manner (see Sec.~\ref{sec:discussion} for a more detailed discussion).

The single-electron Hamiltonian for a single FM qubit is then given by~\cite{Benito2019}
\begin{equation}
    H_\mathrm{1FM} = - \frac{\hbar^2}{2m_\mathrm{e}^\ast} \frac{\partial^2}{\partial x^2}
    + V(x)
    + \frac{g \mu_\mathrm{B} B_x}{2} \sigma_x
    + \frac{g \mu_\mathrm{B} B_z(x)}{2} \sigma_z ,
    \label{H_FM}
\end{equation}
with $\hbar$ the reduced Planck constant, $g$ the g-factor and $\mu_\mathrm{B}$ the Bohr magneton. The double-well confinement potential is denoted by $V(x)$ and $\sigma_x$ and $\sigma_z$ are Pauli matrices. $B_x$ is the homogeneous magnetic field component along the longitudinal axis, defining the spin splitting and qubit quantization axis, while $B_z(x)$ represents the magnetic field gradient profile along the transverse axis (Fig.~\ref{fig:FM_qubits}(b)).

The spatial variation of $B_z(x)$ provides the essential ingredient for EDSR, by transforming electrically-driven displacement of the electron within the double-well into an effective oscillating transverse magnetic drive in the spin frame. Although our framework allows for arbitrary magnetic field profiles $B_z(x)$, in the simulations presented below we adopt a linear gradient,
\begin{equation}
    B_z(x) = b_z x ,
    \label{b_grad}
\end{equation}
motivated by experimentally relevant micromagnet geometries.\cite{DumoulinStuyck2021} Here, $b_z$ quantifies the strength of the field gradient. For completeness, we compare the results obtained from this choice with a step-like gradient profile in Appendix~\ref{app:gradient_comparisions}, which disregards the spatial variation within the wells, making the step-profile similar to the effective low-energy modeling approach.~\cite{Benito2019}

For the double-well confinement, we consider a potential $V(x)$ of the following quartic form:
\begin{equation}
\begin{split}
    V(x) &= V_0(x) + V_\mathrm{d}(x) \\
         &= \frac{\varepsilon_\mathrm{b}}{(d_\mathrm{LR}/2)^4} \left[x^2 - \left(\frac{d_\mathrm{LR}}{2}\right)^2\right]^2 + \frac{\varepsilon_\mathrm{d}}{d_\mathrm{LR}} x.
\end{split}
\label{eq:potential}
\end{equation}
The first term $V_0(x)$ is the symmetric double-well potential where $d_\mathrm{LR}$ is the distance between the minima of the two wells (see Fig.~\ref{fig:FM_qubits}(a)) and $\varepsilon_\mathrm{b}$ is the barrier height (see Fig.~\ref{fig:FM_qubits}(b)). The second term $V_\mathrm{d}(x)$ is a linear detuning potential with $\varepsilon_\mathrm{d}$ quantifying the asymmetry between the two wells (see Fig.~\ref{fig:FM_qubits}(b)).
We note that, for large detuning values, the true barrier height, defined as the value of the potential at the midpoint between the two wells, is not exactly equal to $\varepsilon_\mathrm{b}$, but acquires a detuning-dependent correction. However, for realistic tunnel coupling values (up to $\sim$$10\,\mathrm{GHz}$), this correction is negligible.~\cite{Li2010}

The spectrum of the Hamiltonian is computed by using spinors with the orbital part expanded over one-dimensional harmonic oscillator wavefunctions $\phi_n(x)$ (see Appendix~\ref{app:hermite_polynomials} for details). For the spin degree of freedom, we consider the states $\ket{\uparrow}$ and $\ket{\downarrow}$ denoting the two spin projections along $x$. Thus, our complete basis set is: $\ket{n, \sigma} \equiv \ket{n} \otimes \ket{\sigma}$ ($\sigma = \uparrow, \downarrow$). We restrict our orbital basis set to $n_\mathrm{max} = 50$, with a convergence check provided in Appendix~\ref{app:convergence}.

A crucial modeling choice for the double-well potential when considering this basis set is the choice of the characteristic length of the harmonic oscillator $\ell$ and the center point of the Hermite polynomials (see Eq.~\eqref{HO_basis} in Appendix~\ref{app:hermite_polynomials}). We center the basis around the midpoint of the double-well potential and adjust $\ell$ appropriately to the characteristic length scale of the double-well minima at zero detuning:
\begin{equation}
    \ell \equiv \left( \frac{\hbar^2}{m_\mathrm{e}^\ast \, V''(x)|_{x=\pm d_\mathrm{LR}/2}} \right)^{1/4} = \frac{1}{2} \left( \frac{\hbar^2 d_\mathrm{LR}^2}{2 m_\mathrm{e}^\ast \varepsilon_\mathrm{b}} \right)^{1/4}
\label{eq:def-ell}
\end{equation}
These choices preserve the symmetry of the problem and enable analytic evaluation of matrix elements.~\cite{Shehata2023} An example calculation of the simplified integrals is shown in Appendix~\ref{app:hermite_polynomials}.
This basis thus enables an efficient diagonalization of the single FM qubit Hamiltonian $H_\mathrm{1FM}$ with which we can obtain spatially resolved wavefunctions of the low-energy states (see Fig.~\ref{fig:FM_qubits}(c)).

Unless otherwise stated, we consider the following values for our model parameters to represent FM qubits in a Si-based gate stack: $m_\mathrm{e}^\ast = 0.19\, m_\mathrm{e}$, $g=2$, $B_x = 0.2\,\text{T}$, $b_z = 2\,\text{mT/nm}$, $d_\mathrm{LR} = 100\,\text{nm}$. A detailed list of all the model parameters can be found in Appendix~\ref{app:table}.

\subsection{Two flopping-mode qubits with Coulomb interaction}
\label{subsec:two_FM_model}
We now extend the microscopic framework to two neighboring FM qubits ($\mathrm{FM}_\mathrm{A}$ and $\mathrm{FM}_\mathrm{B}$) in a capacitive-coupling regime with negligible wavefunction overlap (see Fig.~\ref{fig:FM_qubits}(a)). Due to the interplay of Coulomb interaction and magnetic field gradient, an exchange interaction is induced between the two FM qubits,~\cite{Cayao2020} which we compute with a restricted configuration interaction approach.~\cite{Hylleraas1930, Burkard1999, Willmes2025}

We start from the single-electron basis for two spatially separated FM qubits with zero detuning, centered around $x = \pm d_\text{FM}/2$ (see Fig.~\ref{fig:FM_qubits}(a)) with $d_\mathrm{FM}$ the distance between the FM qubits and the signs $+$ and $-$ corresponding to $\mathrm{FM}_\mathrm{B}$ and $\mathrm{FM}_\mathrm{A}$, respectively. The purely orbital single-electron states in position space are obtained as the eigenstates of
\begin{equation}
    h_0^{(\mathrm{A},\mathrm{B})} = - \frac{\hbar^2}{2m_\mathrm{e}^\ast} \frac{\partial^2}{\partial x_{\textsc{a},\textsc{b}}^2}
    + V_0(x_{\textsc{a},\textsc{b}} \pm d_\text{FM}/2),
\end{equation}
For each FM qubit, we retain the two lowest-energy eigenstates ($n = 0, 1$) along with both spin projections along $x$, yielding the eight single-particle states $\ket{\chi}$ with
\begin{equation}
    \ket{\chi} \equiv \ket{\mathrm{A}/\mathrm{B},\, n=0/1,\, \sigma=\uparrow/\downarrow},
    \label{single_part_states}
\end{equation}
where A/B refers to the eigenstate of $\mathrm{FM}_\mathrm{A}$/$\mathrm{FM}_\mathrm{B}$. Restricting to this low-energy sector captures the dominant spin-charge hybridization of each FM qubit relevant for the exchange interaction. The truncation to the two lowest orbital states per qubit ($n = 0, 1$) is justified 
as the higher-orbital states are at significantly higher energies than the relevant energy scales for exchange, and their contribution is strongly suppressed, as can be checked with a multipole expansion (see Appendix~\ref{app:convergence}).

By using the Slater-Condon rules, we lift the single-particle basis to the antisymmetrized two-electron Hilbert space, yielding $\binom{8}{2} = 28$ Slater determinants and corresponding two-particle basis states $\ket{\Psi} \in \{\ket{\chi_1, \chi_2}$.
Here, we focus exclusively on the capacitive-coupling regime, with $d_\mathrm{FM}/d_\mathrm{LR}$ sufficiently large such that electron tunneling between the two FM qubits is negligible.
We therefore eliminate the $2\binom{4}{2} = 12$ Slater determinants in which both electrons occupy the same double-well potential, leaving $16$ states with which we construct the restricted $16 \times 16$ two-FM-qubit Hamiltonian $H_{2\mathrm{FM}}$.

Adding all the other relevant contributions to the two-body Hamiltonian, we can write it as
\begin{align}
    H_\mathrm{2FM} &= H_\mathrm{2FM}^{(0)} + H_\mathrm{d} + H_\mathrm{Z} + H_{\mathrm{grad}} + V_\mathrm{C},
    \label{H_2fm} \\
    H_\mathrm{2FM}^{(0)} &\equiv h_0^{(\mathrm{A})} + h_0^{(\mathrm{B})},
\end{align}
where $H_\mathrm{2FM}^{(0)}$ is the two-particle orbital Hamiltonian with respect to which we constructed our two-electron basis states. The additional terms $H_\mathrm{d}$, $H_\mathrm{Z}$, $H_{\mathrm{grad}}$, and $V_\mathrm{C}$ represent the contributions for the FM detuning, constant magnetic field, magnetic field gradient, and Coulomb interaction, respectively.
The explicit forms of the additional Hamiltonian terms are listed below:
\begin{align}
    H_\mathrm{d} &= h_\mathrm{d}^{(\mathrm{A})} + h_\mathrm{d}^{(\mathrm{B})}, \quad h_\mathrm{d}^{(\mathrm{A,B})} = \mp V_\mathrm{d}( x_{\textsc{a},\textsc{b}} \pm d_\mathrm{FM}/2 ) 
    \label{eq:detuning-A-B} \\
    H_\mathrm{Z} &= \frac{g \mu_\mathrm{B} B_x}{2}\left( \sigma_x^{(\mathrm{A})} + \sigma_x^{(\mathrm{B})} \right), \\
    \begin{split}
        H_{\mathrm{grad}} &= \frac{g \mu_\mathrm{B} b_z}{2} \left[ (x_\textsc{a} + d_\mathrm{FM}/2) \, \sigma_z^{(\textnormal{A})} \right. \\
        &\qquad \qquad \, \left. + (x_\textsc{b} - d_\mathrm{FM}/2) \, \sigma_z^{(\textnormal{B})} \right],
    \end{split} \\
    V_\mathrm{C} &= \frac{e^2}{4\pi\epsilon} \frac{1}{\sqrt{(x_\textsc{a} - x_\textsc{b})^2 + \alpha^2}}.
\end{align}
These terms represent identical FM qubits with the same magnetic field gradient profiles (as shown schematically in Fig.~\ref{fig:FM_qubits}(a)) and detuning towards each other to compensate the Coulomb repulsion (hence the opposite signs for $\mathrm{A}$ and $\mathrm{B}$ in Eq.~\eqref{eq:detuning-A-B}). For the Coulomb interaction, we consider the permittivity in silicon $\epsilon = 11.7\,\epsilon_0$ and parameter $\alpha = \ell$ (see Eq.~\eqref{eq:def-ell}), which effectively captures the transverse directions and regularizes the one-dimensional potential.
A sensitivity check for the exchange calculation with respect to $\alpha$ is provided in Appendix~\ref{app:convergence}.
The default value for $d_\mathrm{FM} = 250\,\textnormal{nm}$.
All these Hamiltonian terms provide matrix elements in the $16\times16$ two-electron subspace, which is then diagonalized to resolve the spectrum.

Note that we do not include the detuning potential for constructing the two-electron basis states of this restricted configuration interaction Hamiltonian. The reason for this is that the Coulomb interaction ($\propto 1/R$) introduces a strong effective detuning of the eigenstates, pushing the lowest-energy solutions to the outermost wells. 
We are interested in the near-symmetric FM operating 
configuration, in which the Coulomb repulsion and the applied detuning 
$\varepsilon_\mathrm{d}$ nearly counterbalance each other, leaving the electrons substantially delocalized over both wells of each FM qubit. 
For this regime, the bonding and antibonding states of the undetuned double-well potential $V_0$ form the natural and physically transparent basis, and allow for a direct comparison with the two-qubit double-dot model of Ref.~\onlinecite{Cayao2020}, which is formulated in this same basis at zero effective detuning. 
The Coulomb interaction and detuning are therefore both treated as perturbations $V_\mathrm{C}$ and 
$H_\mathrm{d}$ acting on this fixed symmetric basis.
The symmetric operating point is identified numerically by searching for the value of 
$\varepsilon_\mathrm{d}$ at which the ground state wavefunction density 
peaks are of equal height in the inner and outer wells of each FM qubit.

\section{Results}
\label{sec:Results}

\subsection{Flopping-mode spectrum and spatial properties}
\label{sec:FM_microscopics}
\begin{figure}
    \centering
    \includegraphics[width=\linewidth]{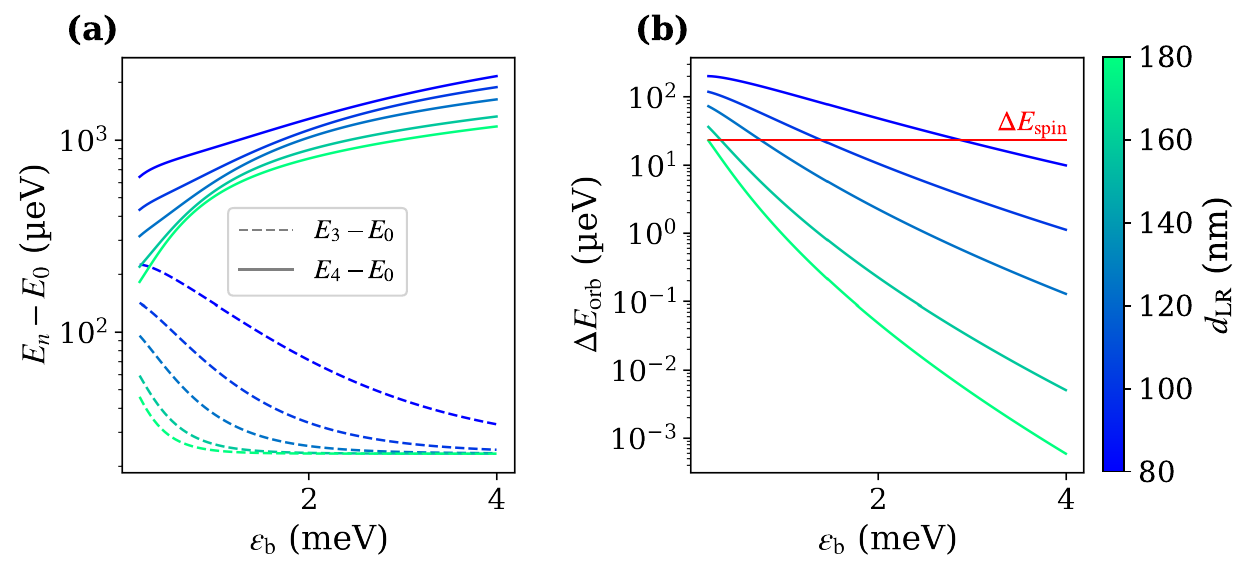}
    \caption{
        \textbf{Orbital level structure of the double-well potential.}
        \textbf{(a)} The energy difference between the ground state ($E_0$) and the third ($E_3$) or fourth ($E_4$) excited state as a function of $\varepsilon_\mathrm{b}$.
        \textbf{(b)} The orbital splitting as a function of $\varepsilon_\mathrm{b}$ (with $\varepsilon_\mathrm{d} = 0$), with a spin splitting corresponding to $|B_x|=0.2\,\text{T}$ indicated by the red horizontal line.
        The results in (a) and (b) are shown for different values of $d_\mathrm{LR}$, as indicated by the color scale, and obtained with $b_z=0$.
    }
    \label{fig:DWP_orbital_structure}
\end{figure}

First, we resolve and analyze the spectrum of $H_\mathrm{1FM}$ without a magnetic field gradient. As expected for a double-well potential, we naturally obtain four low-energy states ($E_{0\leq n \leq 3}$) that are well separated from the higher-energy states ($E_{n \geq 4}$), as can be seen in Fig.~\ref{fig:DWP_orbital_structure}(a). Without detuning, $\epsilon_d=0$, the four lowest-energy states correspond to symmetric (bonding) and antisymmetric (antibonding) orbital states, each with two spin polarizations. The two orbital states have an energy difference between them, which we refer to as \emph{orbital splitting} $\Delta E_\mathrm{orb}$, controlled by $\epsilon_b$ (see Fig.~\ref{fig:DWP_orbital_structure}(b)). For a detuned double-well potential, the orbital eigenstates become asymmetrically delocalized over the left and right quantum well. Apart from the orbital splitting there is the energy difference between the two spin polarizations due to the constant magnetic field, which we refer to as \emph{spin splitting} $\Delta E_\mathrm{spin}$.

\begin{figure}[tb]
    \includegraphics[width=\linewidth]{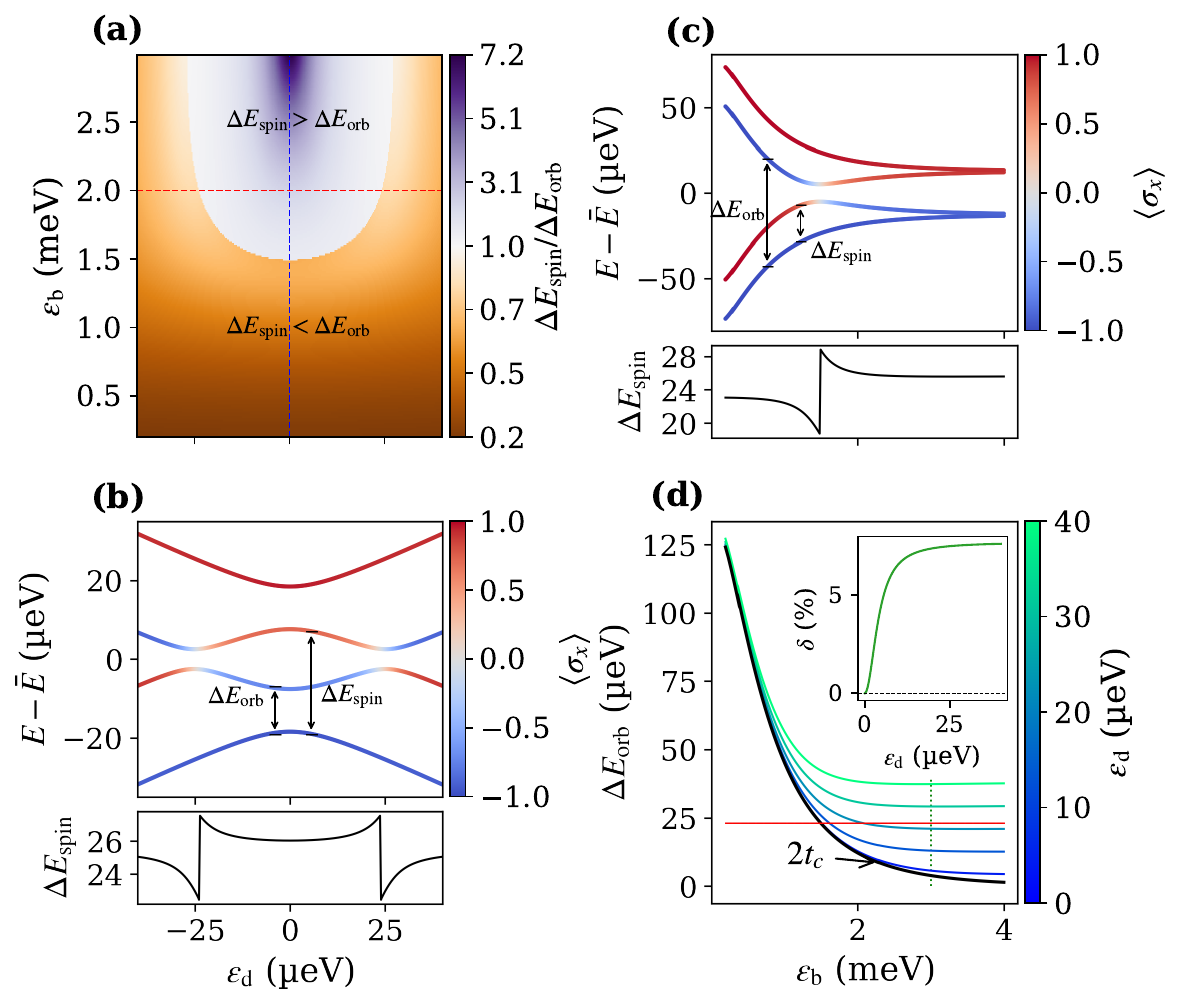}
    \caption{
        \textbf{Flopping-mode qubit spectrum}.
        \textbf{(a)} The ratio of spin and orbital splitting as a function of $\varepsilon_\mathrm{d}$ and $\varepsilon_\mathrm{b}$, with the regimes where spin or orbital splitting dominates indicated.
        \textbf{(b),(c)} The energy of the four lowest-energy eigenstates (relative to their average) and spin splitting as a function of \textbf{(b)} $\varepsilon_\mathrm{d}$ with $\varepsilon_\mathrm{b} = 2\,\textnormal{meV}$ and \textbf{(c)} $\varepsilon_\mathrm{b}$ with $\varepsilon_\mathrm{d} = 0$. The color scale indicates the spin polarization of the states.  
        \textbf{(d)} The orbital splitting as a function of $\varepsilon_\mathrm{b}$ for different detuning values, as indicated by the color scale. The spin splitting for $|B_x| = 0.2\,\textnormal{T}$ is indicated by a red horizontal line for reference.
        The inset shows the difference in orbital splitting compared to the low-energy model expression, $\delta \equiv [\Delta E_\mathrm{orb} - (4 t_\mathrm{c}^2 + \varepsilon_\mathrm{d}^2)^{1/2}]/\Delta E_\mathrm{orb}$, as a function of $\varepsilon_\mathrm{d}$ at $\varepsilon_\mathrm{b}=3\,\text{meV}$ (indicated by a vertical line in the main panel).
    }
    \label{fig:FM_qubit_spectrum}
\end{figure}

With a magnetic field gradient, there is spin-charge hybridization such that the orbital and spin degrees of freedom of the low-energy states get mixed. An example of the resulting wavefunctions of the low-energy spectrum of $H_\mathrm{1FM}$ with detuning and magnetic fields (constant and gradient) is shown in Fig.~\ref{fig:FM_qubits}(c).
In Fig.~\ref{fig:FM_qubit_spectrum}, we evaluate $\Delta E_\mathrm{spin}$ and $\Delta E_\mathrm{orb}$ as a function of the two primary electrostatic control parameters of an FM qubit: $\varepsilon_\mathrm{b}$, which controls the tunnel coupling between the two wells, and $\varepsilon_\mathrm{d}$, which controls the asymmetry between the two wells (see Fig.~\ref{fig:FM_qubits}). We obtain different regimes where either the spin splitting or the orbital splitting dominates (see Fig.~\ref{fig:FM_qubit_spectrum}(a)).~\cite{Benito2019} Due to spin-charge hybridization, there is an avoided crossing appearing between the spin and charge excitations at the phase boundary between the two regimes (see Figs.~\ref{fig:FM_qubit_spectrum}(b) and (c)). Note that, in the presence of spin-charge hybridization, we generally consider $\Delta E_\mathrm{spin}$ ($\Delta E_\mathrm{orb}$) to correspond to the energy difference between the ground state and the next excited state that differs more from the ground state in spin polarization (orbital structure) than in orbital structure (spin polarization). Further note that the accessible windows of spin- and orbital-splitting-dominated regimes as a function of  $\varepsilon_\mathrm{b}$ are naturally constrained by the interplay of magnetic field strength and double-well separation in our microscopic modeling framework (see Fig.~\ref{fig:DWP_orbital_structure}(b)).

A common approach for modeling FM qubits further abstracts the low-energy states of the double-well potential into a double-dot model with spin degree of freedom (see Appendix~\ref{app:low_energy_models}), with effective parameters for detuning and tunnel coupling ($t_c \equiv \Delta E_\mathrm{orb}/2$, with $\Delta E_\mathrm{orb}$ obtained for zero detuning, see Fig.~\ref{fig:FM_qubit_spectrum}(d)) between the two dots. With such a low-energy model, the detailed spatial properties of the low-energy states, which describe the delocalization of the states across the double-well potential and the connection to microscopic device parameters, are lost. This can result in a misestimate of the orbital splitting of a detuned FM qubit, for example (see inset of Fig.~\ref{fig:FM_qubit_spectrum}(d)).

\begin{figure}[tb]
    \includegraphics[width=\linewidth]{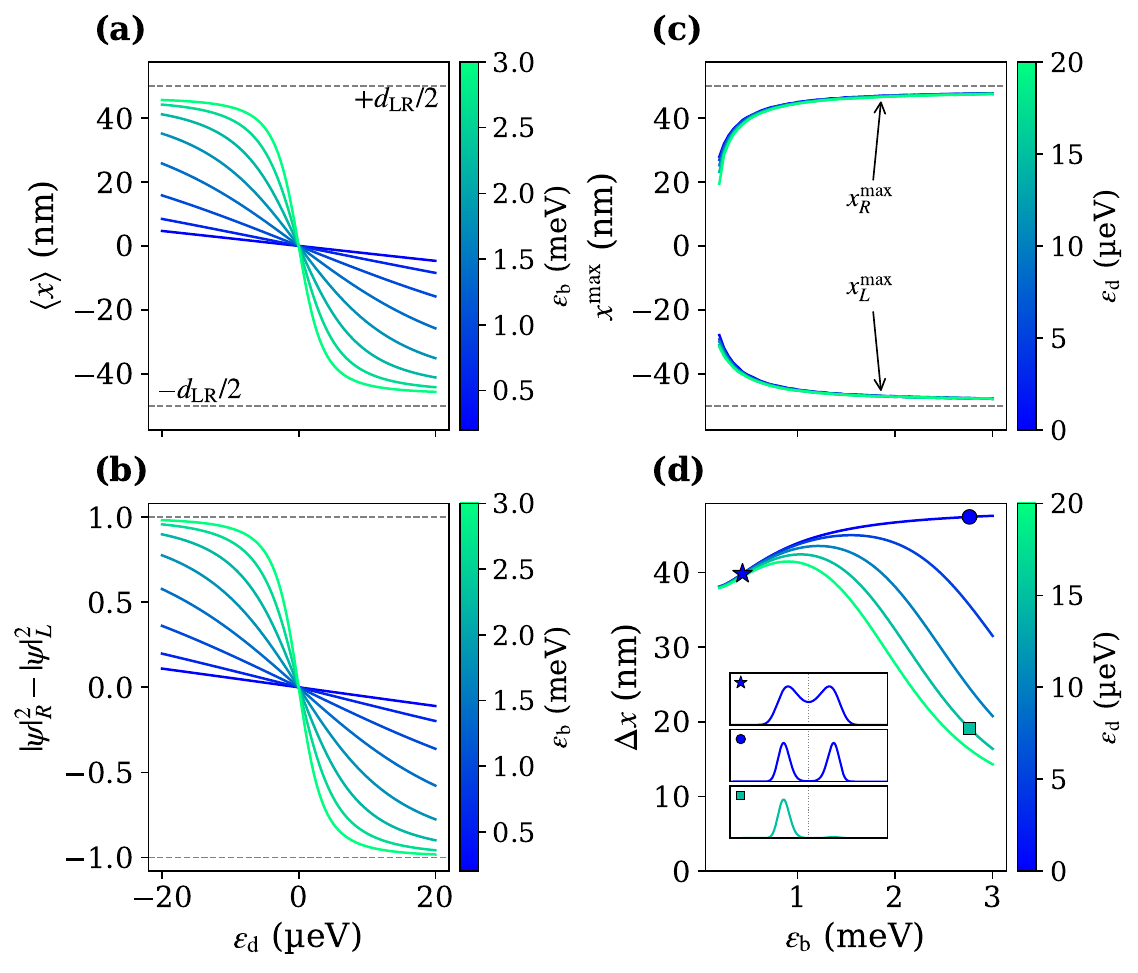}
    \caption{
        \textbf{Spatial properties of the flopping-mode ground state}.
        \textbf{(a)},\textbf{(b)} The \textbf{(a)} mean electron position $\langle x \rangle$ and \textbf{(b)} charge asymmetry $|\psi|_\mathrm{R}^2 - |\psi|_\mathrm{L}^2$ as a function of $\varepsilon_\mathrm{d}$ for different values of $\varepsilon_\mathrm{b}$, as indicated by the color scale. Here,  $|\psi|_\mathrm{L}^2$ and $|\psi|_\mathrm{R}^2$ denote the integrated wave function density of the left ($x<0$) and right ($x>0$) well, respectively.
        \textbf{(c)},\textbf{(d)} The \textbf{(c)} peak position $x^\mathrm{max}$ of the left and right wavefunction maxima in the double well (see Fig.~\ref{fig:FM_qubits}(c)) and \textbf{(d)} spread of the wavefunction $\Delta x \equiv (\langle x^2 \rangle - \langle x \rangle^2)^{1/2}$ as a function of $\varepsilon_\mathrm{b}$ for different values of $\varepsilon_\mathrm{d}$, as indicated by the color scale. The inset shows ground state wavefunctions at the indicated points, clearly showing where single/double-dot physics dominates.
    }
    \label{fig:FM_spatialprops}
\end{figure}

The spatial properties directly influence the effective dipole moment of an FM qubit as well as its Coulomb interaction strength with a neighboring FM qubit. They are thus relevant for a detailed assessment of the FM qubit performance with respect to EDSR-based single-qubit and exchange-based two-qubit control, which follows in the subsections below.
The impact of $\varepsilon_\mathrm{d}$ or $\varepsilon_\mathrm{b}$ on the spatial properties of the low-energy states (in particular the ground state) is presented in Fig.~\ref{fig:FM_spatialprops}. Figs.~\ref{fig:FM_spatialprops}(a) and (b) show that, at zero detuning, the electron sits symmetrically between the two wells ($\langle x \rangle = 0$, $|\psi|_L^2 - |\psi|_R^2 = 0$). Increasing $\varepsilon_\mathrm{d}$ shifts the electron towards one well, and increasing $\varepsilon_\mathrm{b}$ sharpens the transition to single-well localization as the tunnel coupling $t_\mathrm{c}$ and wave function density in between the two wells are suppressed. At low $\varepsilon_\mathrm{b}$, the wavefunction density peaks deviate significantly from the double-well minima [Fig.~\ref{fig:FM_spatialprops}(c)], reflecting the incomplete separation of the two wells. The nonmonotonic variation of $\Delta x$ with $\varepsilon_\mathrm{b}$ at nonzero detuning [Fig.~\ref{fig:FM_spatialprops}(d)] marks the crossover between single- and double-dot behaviour of the FM qubit as the barrier rises.

\subsection{Single-qubit control: electric dipole spin resonance}
\label{sec:EDSR}
In this section, we investigate single-qubit control. In particular, we consider the EDSR-driven dynamics of a single FM qubit.
For this, we add a time-dependent driving term of the form $H_\mathrm{EDSR}(t) = e E_0 x \sin(\omega_\mathrm{EDSR} t)$ to $H_\mathrm{1FM}$, where $E_0$ is the amplitude of the applied electric field and $\omega_\mathrm{EDSR}$ the (angular) drive frequency.
We consider a drive frequency equal to $\Delta E_\mathrm{spin}/\hbar$ to drive transitions between the ground state and the first excited state with opposite spin polarization (separated from the ground state with energy $\Delta E_\mathrm{spin}$), which we consider as the computational basis states of our FM qubit.
Practically, such a drive can be implemented by applying an ac voltage between the two plunger gates of a single FM qubit (see Fig.~\ref{fig:FM_qubits} (a)).
This drive couples directly to the electron position operator $x$, which induces transitions between orbital states.
Due to the transverse magnetic field gradient, the orbital part couples to the spin such that spin rotations are induced by the drive.

\begin{figure*}[tb]
    \centering
    \includegraphics[width=\linewidth]{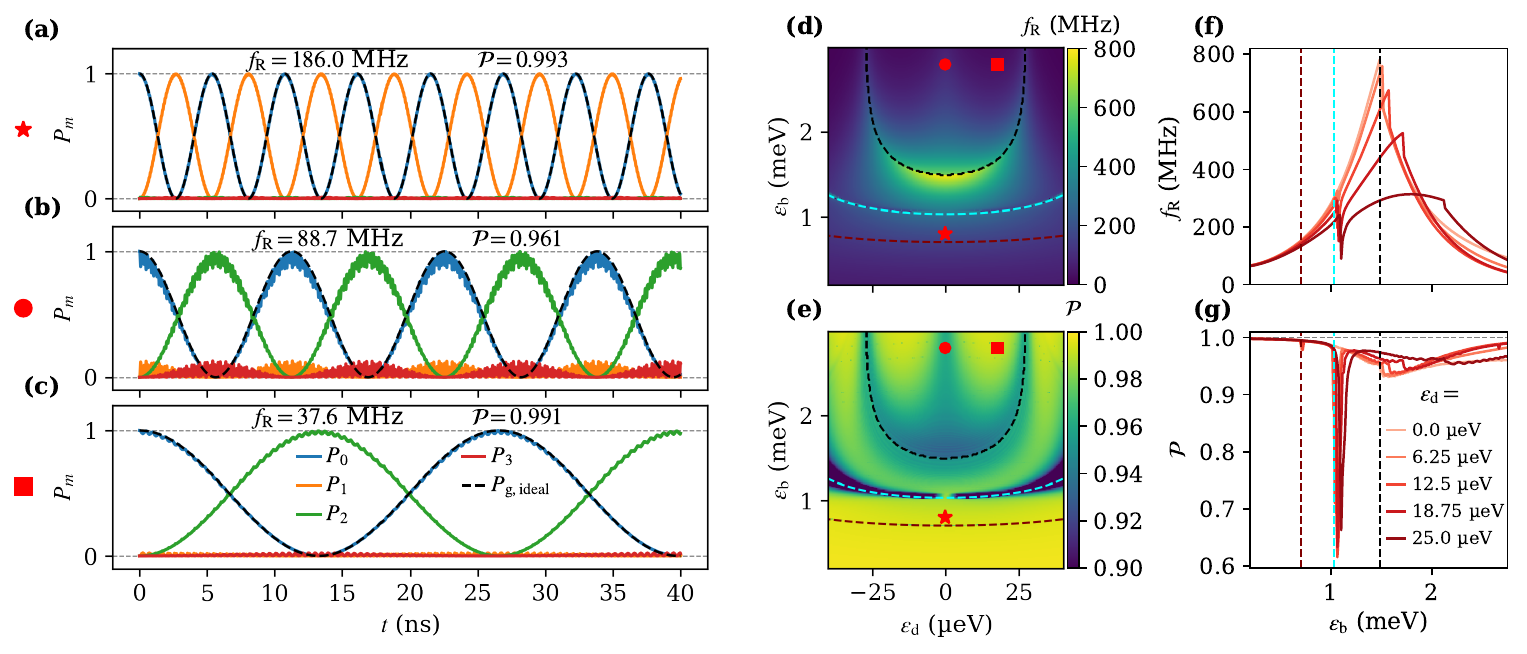}
    \caption{
        \textbf{Rabi oscillations of an EDSR-driven flopping-mode qubit.}
        \textbf{(a)}-\textbf{(c)} The probabilities of the four lowest-energy states of a flopping-mode qubit as a function of time during an EDSR drive in three different parameter configurations for detuning $\epsilon_\mathrm{d}$ and barrier height $\epsilon_\mathrm{b}$, corresponding to \textbf{(a)} $(\varepsilon_\mathrm{d}, \varepsilon_\mathrm{b}) = (0, 0.8\,\text{meV})$, \textbf{(b)} $(0, 2.8\,\text{meV})$, and \textbf{(c)} $(18\,\text{\textmu eV}, 2.8\,\text{meV})$. The black dashed line indicates the ideal (single-frequency) Rabi oscillation pattern of the ground state, as extracted from a Fourier analysis (see Appendix~\ref{app:additional-spectral-properties}).
        \textbf{(d)},\textbf{(e)} The \textbf{(d)} Rabi frequency $f_\mathrm{R}$ and \textbf{(e)} spectral purity $\mathcal{P}$ (see Eq.~\eqref{eq:spectral-purity}) as a function of $\epsilon_\mathrm{d}$ and $\epsilon_\mathrm{b}$, with the examples of (a)-(c) indicated with matching symbols. 
        The black dashed contour line indicates the spin-to-orbital splitting-dominated crossover (see Fig.~\ref{fig:FM_qubit_spectrum}(a)). The cyan and marron dashed contour lines indicate where the orbital splitting is twice and thrice the spin splitting, respectively.
        \textbf{(f)}, \textbf{(g)} Vertical line cuts of \textbf{(f)} $f_\mathrm{R}$ and \textbf{(g)} $\mathcal{P}$ from (d) and (e) as a function of $\varepsilon_\mathrm{b}$ for different values of $\varepsilon_\mathrm{d}$.
        The vertical dashed lines indicate the values of $\epsilon_\mathrm{b}$ where the vertical line cuts cross the contour lines in (d),(e).
    }
    \label{fig:FM_EDSR_1}
\end{figure*}

\begin{figure}[htb]
    \centering
    \includegraphics[width=\linewidth]{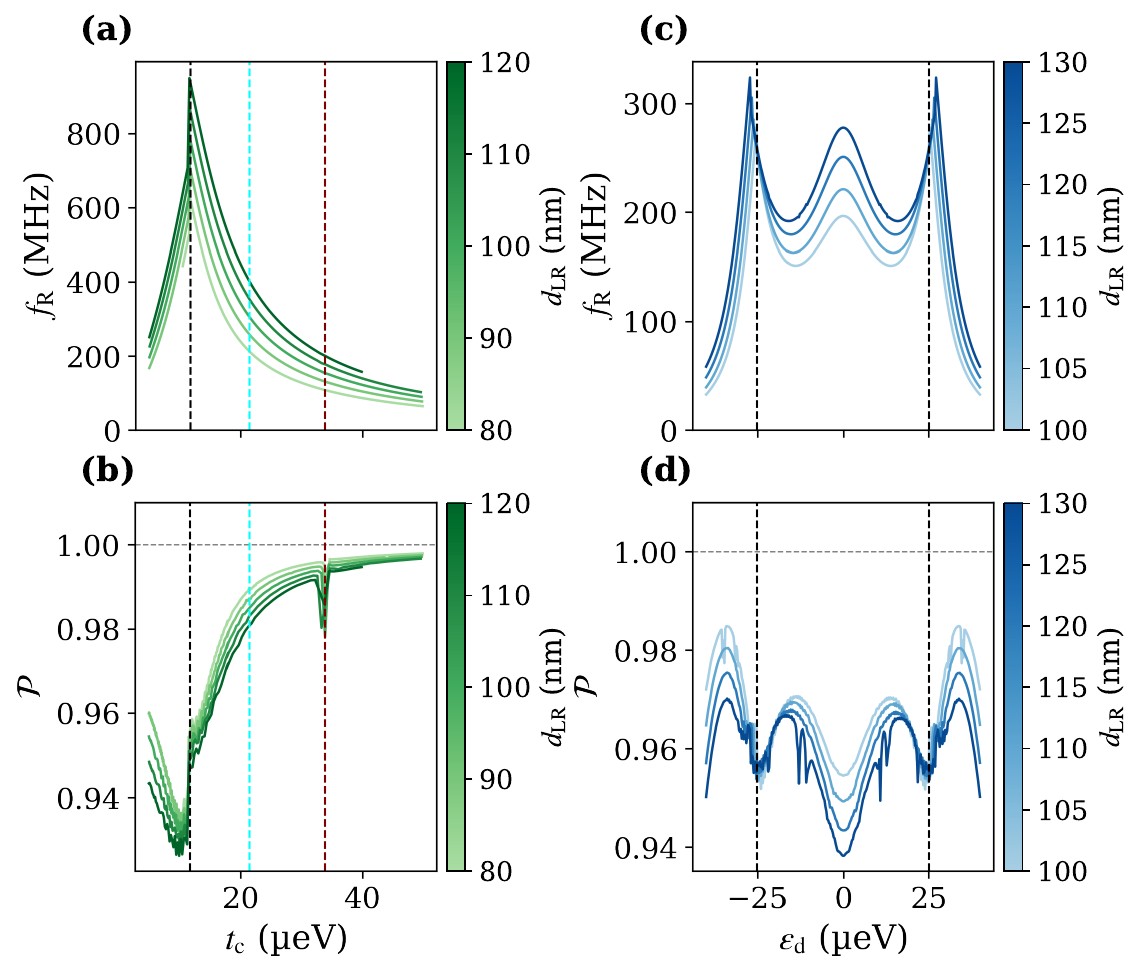}
    \caption{
        \textbf{Dependence of Rabi frequency and spectral purity on double-well separation}.
        \textbf{(a)},\textbf{(b)} Values of \textbf{(a)} $f_\mathrm{R}$ and \textbf{(b)} $\mathcal{P}$ as a function of $t_\mathrm{c} \equiv \Delta E_\mathrm{orb}/2$ (see Fig.~\ref{fig:DWP_orbital_structure}(b)) with $\varepsilon_\mathrm{d} = 0$ for different double-well separations $d_\mathrm{LR}$.
        \textbf{(c)},\textbf{(d)} Values of \textbf{(c)} $f_\mathrm{R}$ and \textbf{(d)} $\mathcal{P}$ as a function of $\varepsilon_\mathrm{d}$ for different double-well separations $d_\mathrm{LR}$ with fixed $t_\mathrm{c} = 5\,\text{\textmu eV}$ (by appropriately adjusting $\varepsilon_\mathrm{b}$).
        The vertical black dashed line indicates the spin-to-orbital splitting-dominated crossover (see Fig.~\ref{fig:FM_qubit_spectrum}(a)).
        The cyan and marron dashed lines in (a),(b) indicate where the orbital splitting is twice and thrice the spin splitting, respectively.
    }
    \label{fig:FM_EDSR_2}
\end{figure}

To analyze the EDSR drive, we numerically evaluate the time evolution operator using a standard Trotter decomposition of the time evolution operator, i.e. the time-ordered exponential:
\begin{equation}
    U(t) \equiv \mathcal{T}\exp\left[-\frac{i}{\hbar}\int_0^t H(t') dt' \right],
\end{equation}
with $H(t) = H_\mathrm{1FM} + H_\mathrm{EDSR}(t)$.
The Hamiltonian at each time step is assembled from matrix elements evaluated analytically within the Hermite polynomial basis (Appendix~\ref{app:hermite_polynomials}).

We apply the time evolution operator to the FM qubit initialized in the ground state and evaluate the time dependent probabilities $P_m(t)$ in a given eigenstate $m$ of $H_\mathrm{1FM}$, with $m=0,1,2,\ldots$ labeling the different eigenstates sorted by their energy in ascending order (see Fig.~\ref{fig:FM_qubits}(c)). For the relevant range of model parameters of $H_\mathrm{1FM}$ for an FM qubit, the probabilities remain within the four-dimensional low-energy subspace such that $P_{m\geq4} \approx 0$ and higher-energy states can safely be neglected. The time evolution of the probabilities of these low-energy states during the EDSR drive is shown for three different combinations of $(\varepsilon_\mathrm{d},\varepsilon_\mathrm{b})$ in Figs.~\ref{fig:FM_EDSR_1}(a)-(c). From the resulting probability oscillations, we extract the Rabi frequency $f_\mathrm{R}$ by performing a Fourier analysis on $P_0(t)$ and identifying the dominant frequency in the oscillation (shown as a black dashed line), which describes the single-frequency Rabi pattern (which we call the ideal Rabi oscillation pattern) through $P_{\mathrm{ideal}}(t) = \cos^2(\pi f_\mathrm{R} t)$.
It is clear that the Rabi frequency strongly depends on this choice of parameters $(\varepsilon_\mathrm{d},\varepsilon_\mathrm{b})$.

As seen from the examples, the oscillations do not always nicely follow the ideal Rabi oscillation pattern. Significant higher-frequency corrections can be observed for certain combinations of barrier height and detuning (see Appendix~\ref{app:additional-spectral-properties} for a detailed Fourier analysis).
Due to the presence of orbital states in the low-energy subspace of the FM qubit, the EDSR drive can induce coherent leakage into states outside the computational subspace, even in the absence of environmental noise. This leakage affects the quality of the Rabi oscillations and may render difficult the implementation of quantum gates (e.g. a $\pi/2$ rotation) with high fidelity.
To study the impact of this leakage effect on the Rabi oscillations, we compare the oscillating ground-state population $P_{0}(t)$, as obtained with our model, with the ideal Rabi oscillation profile $P_{\mathrm{ideal}}(t) = \cos^2(\pi f_\mathrm{R} t)$.
We can quantify this impact with the quantity $\mathcal{P}$, which we refer to as the \emph{spectral purity} of the Rabi oscillation pattern, defined as:

\begin{equation}
    \mathcal{P} \equiv
    1 - \frac{1}{(2/\pi)}
    \left\langle
    \left| P_\mathrm{0}(t) - P_{\mathrm{ideal}}(t) \right|
    \right\rangle_{t \in [0, \Delta t]},
\label{eq:spectral-purity}
\end{equation}
where the average is taken over the time interval $\Delta t$, which we set equal to two Rabi periods.
With this definition, $\mathcal{P}=1$ corresponds to an ideal single-frequency Rabi oscillation pattern, while a reduction in $\mathcal{P}$ indicates a deviation from the ideal two-level dynamics.
The spectral purity is normalized so that a flat $P_0(t) = 0.5$ yields $\mathcal{P} = 0$. 
Note that the spectral purity can also be quantified through the spectral peaks of the oscillation pattern in the frequency domain. This yields similar results, as shown with an example in Appendix~\ref{app:additional-spectral-properties}.
We emphasize that $\mathcal{P}$ quantifies the intrinsic dynamics of the driven qubit alone, evaluated in a fully coherent, noise-free simulation. 
The deviations from an ideal single-frequency Rabi oscillation that reduce $\mathcal{P}$ below unity are not associated with environmental noise, but reflect genuine coherent leakage into higher orbital states induced by the EDSR drive itself, and are therefore an intrinsic feature of the FM qubit's multi-level structure. $\mathcal{P}$ thus captures a distinct, coherent source of gate imperfection that complements noise-driven fidelity metrics. 
Physically, a reduction in $\mathcal{P}$ manifests experimentally as a distortion of the Rabi chevron pattern and directly increases the precision required in the pulse duration to implement a high-fidelity gate (e.g.\ a $\pi/2$ rotation). 
The additional frequency content present in a low-purity oscillation causes the ground-state population to deviate more strongly from the ideal $\cos^2$ envelope, so a given timing uncertainty will produce a larger population error than under a high-quality single-frequency oscillation.

In Fig.~\ref{fig:FM_EDSR_1}(d), we present the extracted Rabi frequency as a function of detuning and barrier height across the spin- and orbital-splitting-dominated regimes of the FM qubit, with linecuts for different detuning values in (f). Together, they show that the highest Rabi frequencies are obtained near the transition between the two regimes, in particular when the detuning is relatively small. This is expected because spin-charge hybridization is maximal and the wavefunctions are maximally delocalized over the double-well in this regime, maximizing the effective electric dipole moment of the electron and the coupling of the EDSR drive to the spin.

Similar to the Rabi frequency, we present the spectral purity of the Rabi oscillations as a function of detuning and barrier height across the spin- and orbital-splitting-dominated regimes of the FM qubit in Fig.~\ref{fig:FM_EDSR_1}(e) with linecuts shown in (g).
The conditions for a high spectral purity are noticeably different from those that yield a high Rabi frequency. The spectral purity is generally high in the regime where the spin splitting is smaller than orbital splitting, independent of detuning. In this regime, the two spin states of the computational basis have the lowest energy in the system, and leakage to higher-energy orbital states is generally suppressed. This holds generally, except when the orbital splitting is a multiple of the spin splitting. In this case, the higher harmonics of the drive frequency come into resonance with the orbital transition, enabling leakage into higher-energy orbital states through multi-photon excitation processes (the cyan and maroon contour lines indicate these resonance conditions). As a consequence, dips in Rabi frequency or spectral purity can be observed. Note that the two-photon excitation process, which is the multi-photon process that affects the spectral purity most strongly, has no impact near zero detuning due to the odd-even parity symmetry of the orbital states.
When orbital splitting is smaller than spin splitting, the spectral purity is maximal in between the spin-to-orbital crossover and the zero-detuning regime. This is the regime in which spin-charge hybridization or charge delocalization is minimal and, correspondingly, the Rabi frequency is relatively low. 

Fig.~\ref{fig:FM_EDSR_2} shows how the Rabi frequency and spectral purity evolve with double-well separation across a range of barrier height and detuning values, providing an assessment of single-qubit control directly in terms of the device geometry.
In general, $f_\mathrm{R}$ increases with increasing $d_\mathrm{LR}$.
This is because the larger double-well separation amplifies the spatial extent of the low-energy states across the magnetic field gradient, thus improving the EDSR drive efficiency.
However, the spectral purity is also generally reduced, as increased spin-charge hybridization also enhances leakage outside of the computational basis.
Note that, even though the barrier heights are adjusted appropriately to plot $f_\mathrm{R}$ and $\mathcal{P}$ as a function of $t_\mathrm{c}$ for every double-well separation in Fig.~\ref{fig:FM_EDSR_2}, the curves remain different. Consequently, the different device geometries cannot be described by the same effective double-dot model (Appendix~\ref{app:low_energy_models}).

When considering the combined results of Rabi frequency and spectral purity of the corresponding oscillations, we end up with a natural tradeoff between the optimal conditions for \emph{fast} and \emph{clean} (single-frequency) Rabi oscillations for the EDSR-driven FM qubit. This tradeoff applies across different double-well separations, with the optimization with respect to Rabi frequency and spectral purity generally pushing the FM qubit in opposite directions in the $(\varepsilon_\mathrm{d}, \varepsilon_\mathrm{b})$ parameter space.

\subsection{Two-qubit control: capacitive exchange coupling}
\label{sec:exchange}
While the conventional scheme for the exchange coupling of spin qubits relies on direct wavefunction overlap, the spin-charge hybridization in FM qubits due to the magnetic field gradient profile enables exchange interaction through capacitive coupling.
In Ref.~\onlinecite{Cayao2020}, the following expression was derived for the singlet-triplet energy splitting $J_\mathrm{DD}$, based on a double-dot model for the FM qubits in the setup under consideration here (i.e. two identical neighboring FM qubits, see Fig.~\ref{fig:FM_qubits}(a) and Appendix~\ref{app:low_energy_models}):
\begin{equation}
    J_\mathrm{DD} = 2 \left| \frac{4 E_z^2 t_\mathrm{c}^2 \kappa}{(E_x^2 - 4 t_\mathrm{c}^2)^2 - 4 E_x^2 \kappa^2} \right|,
    \label{eq:J_analytic}
\end{equation}
with $E_z = g \mu_\mathrm{B} b_z d_\mathrm{LR}/2$ the effective Zeeman splitting between the two double-well potential minima due to the magnetic field gradient, $E_x = g \mu_\mathrm{B} B_x$ the Zeeman splitting due to the constant magnetic field, $t_\mathrm{c}$ the tunnel coupling within each qubit (considered identical), and $\kappa$ an effective capacitive-coupling parameter for the Coulomb interaction between the two qubits (see Appendix~\ref{app:low_energy_models} for details). The expression shows that the exchange strength can be controlled by the effective magnetic field strength of the gradient profile, as well as the internal FM qubit characteristics (tunnel coupling $t_\mathrm{c}$). Note that this expression assumes zero \emph{effective} detuning (see below). Note also that this analytical expression is perturbative in $\kappa$ and formally valid only when $\kappa \ll \kappa_\mathrm{div} \equiv |E_x^2 - 4 t_\mathrm{c}^2|/(2 E_x)$, with the expression diverging at $\kappa = \kappa_\mathrm{div}$. 

\begin{figure}[tb]
    \includegraphics[width=\linewidth]{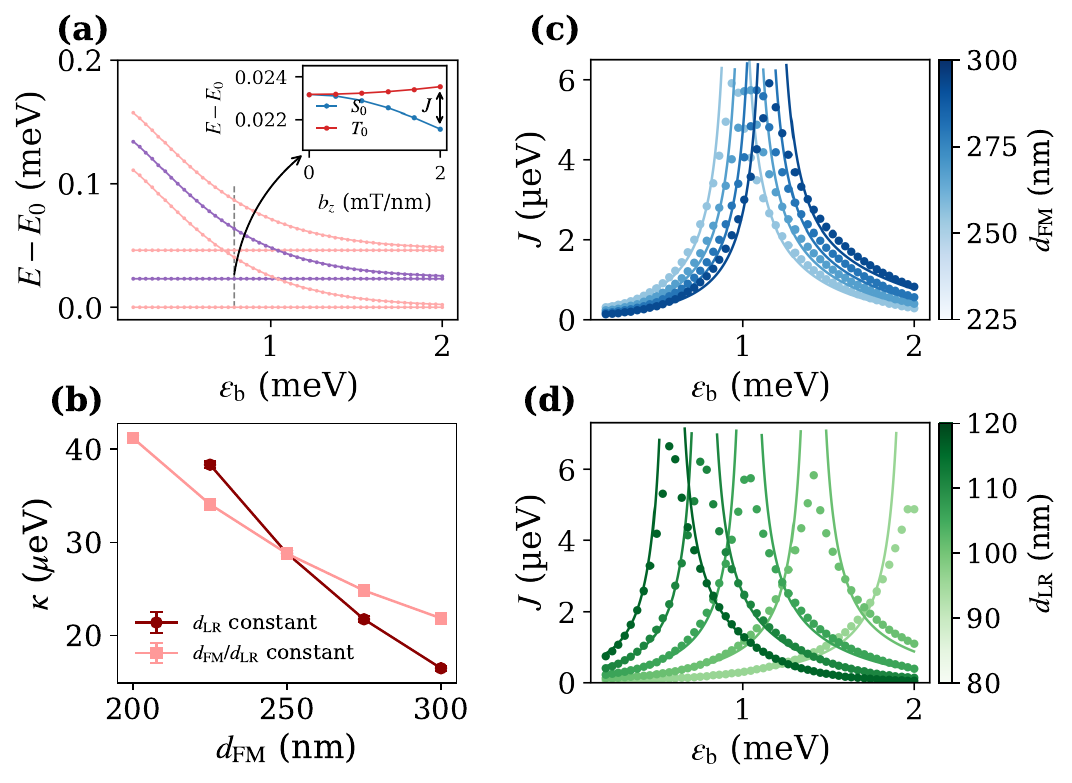}
    \caption{
        \textbf{Exchange strength of two capacitively-coupled flopping-mode qubits}.
        \textbf{(a)} Eigenenergies of the first eight states of $H_\mathrm{2FM}$ (see Eq.~\eqref{H_2fm}) relative to the ground state as a function of $\varepsilon_\mathrm{b}$ at $b_z = 0$. The purple lines indicate the different singlet (S0)-triplet (T0) pairs. The inset shows the singlet-triplet splitting of the lowest-energy pair as a function of $b_z$ (with $\varepsilon_\mathrm{b} = 0.8\,\text{meV}$).
        \textbf{(b)} Effective capacitive-coupling parameter $\kappa$ (see Eq.~\eqref{eq:J_analytic}) as a function of $d_\mathrm{FM}$, keeping either $d_\mathrm{LR} = 100\,\text{nm}$ constant or $d_\mathrm{FM}/d_\mathrm{LR} = 5/2$ fixed.
        \textbf{(c)},\textbf{(d)} Exchange strength $J$ as a function of $\varepsilon_\mathrm{b}$ for \textbf{(c)} different qubit separations $d_\mathrm{FM}$ and \textbf{(d)} for different double-well separations $d_\mathrm{LR}$ of the two qubits, keeping $d_\mathrm{FM}/d_\mathrm{LR} = 5/2$ fixed. 
        In both panels, dotted points are extracted from eigensolutions of $H_\mathrm{2FM}$ (see eq.~\ref{H_2fm}) and solid lines are fits to Eq.~\eqref{eq:J_analytic}.
    }
    \label{fig:FM_exchange}
\end{figure}

Here, we calculate the exchange strength based on the microscopic model for two neighboring FM qubits of which the electrons interact directly via the Coulomb interaction (without spatial wavefunction overlap). Through diagonalization of $H_\mathrm{2FM}$, we extract the energy difference between the lowest-energy singlet-triplet pair in the system (see Fig.~\ref{fig:FM_exchange}(a)), from which we extract the \emph{capacitive exchange strength} $J \equiv |E_T - E_S|$, with $E_S$ ($E_T$) the energy of the singlet (triplet). This can be compared to the double-dot expression $J_\mathrm{DD}$ from Eq.~\eqref{eq:J_analytic}.

Throughout this analysis, the exchange interaction is evaluated at zero \emph{effective} detuning of each FM qubit, where the electrostatic detuning compensates the Coulomb repulsion and restores maximally delocalized low-energy eigenstates across the double-well potential. The corresponding value of $\varepsilon_\mathrm{d}$ is determined by searching through a range of $\varepsilon_\mathrm{d}$ values for the value that evenly distributes the wavefunction density of the ground state of $H_\mathrm{2FM}$ over the double well of each FM qubit.

By fitting the exchange strength to the expression in Eq.~\eqref{eq:J_analytic}, we can extract a microscopically derived value for $\kappa$. We consider two different scenarios for our analysis: (i) keeping the double-well separation fixed ($d_\mathrm{LR}$ constant) or (ii) scaling it proportional to the inter-qubit separation ($d_\mathrm{LR} \propto d_\mathrm{FM}$). The corresponding exchange strength of the two scenarios is shown as a function of barrier height in Figs.~\ref{fig:FM_exchange}(c) and (d), respectively, together with the fits (for which $\varepsilon_\mathrm{b}$ is converted to $t_\mathrm{c}$ as shown in Fig.~\ref{fig:DWP_orbital_structure}(b)). 
The key finding is that $J$ peaks at an intermediate barrier height and that increasing $d_\mathrm{LR}$ proportionally with $d_\mathrm{FM}$ partially compensates the weakening of the Coulomb interaction at larger inter-qubit separations.
The fits nicely capture the microscopically derived values, except near the divergence at $\kappa = \kappa_\mathrm{div}$, where the effective double-dot expression $J_\mathrm{DD}$ diverges unphysically. The microscopically obtained values of $J$, by contrast, remain regular and finite in this regime. This is a qualitatively important difference: the divergence in the double-dot model is an artifact of the perturbative treatment of $\kappa$, whereas the microscopic model captures the full Coulomb interaction without such an approximation, naturally regularizing the exchange coupling. This highlights one of the key advantages of the present framework over reduced effective models. 

The results of the two scenarios described above are compared in Fig.~\ref{fig:FM_exchange}(b) as a function of the separation between the two FM qubits.Although both scenarios yield a monotonic decrease of $\kappa$ with inter-qubit separation due to the weakening Coulomb interaction, the decrease can be mitigated by scaling the intra-qubit double-well separation proportionally with the inter-qubit separation.

Note that, unlike in Eq.~\eqref{eq:J_analytic}, $\kappa$ is not a truly independent parameter in our microscopic model. A change of $\varepsilon_\mathrm{b}$ (or, correspondingly, $t_\mathrm{c}$) modifies the Coulomb integrals and thereby changes the effective capacitive-coupling parameter. Nonetheless, we consider it an independent fitting parameter, which should be interpreted as an effective capacitive-coupling strength that is averaged over the sampled range of $\varepsilon_\mathrm{b} = 0.5$-$2$~meV. The fitted values of $\kappa \approx 20$-$40~\mu$eV are therefore specific to the device geometries studied ($d_\mathrm{LR} = 100$~nm, $d_\mathrm{FM} = 225$-$300$~nm) and should not be extrapolated to significantly different parameter regimes, in particular to barrier heights beyond $\varepsilon_\mathrm{b} = 2$~meV where the wavefunction becomes strongly localized, or to well separations and inter-qubit distances substantially outside the ranges studied, where the Coulomb matrix elements and wavefunction profiles may differ considerably from those underlying the fits.

\section{Discussion and Outlook}
\label{sec:discussion}

In this section, we discuss two aspects of our methodology that merit some consideration: the validity of the one-dimensional approximation and the sensitivity of the results to the assumed field profile.

The one-dimensional approximation is expected to be reliable in the regime where the transverse confinement energy is large compared to the relevant longitudinal energy scales, namely the tunnel coupling, detuning, and Zeeman splitting. 
In typical gate-defined silicon or germanium quantum dots, the transverse confinement energy is on the order of several meV,~\cite{Burkard2023} which is larger than or comparable to the barrier heights considered here, and well above the tunnel coupling and Zeeman energy scales. 
Effects not captured by the one-dimensional description include transverse orbital excitations, interface roughness, electrostatic disorder, and, in silicon-based platforms, valley degrees of freedom.~\cite{Losert2025} 
Valley splitting in particular can affect the spin relaxation time and modify the effective spin-charge hybridization when the valley splitting is comparable to the orbital splitting. 
These effects are beyond the scope of the present work and are discussed further in the context of future extensions.

The quartic double-well potential and linear gradient profile studied here represent a particular demonstration of our methodology. Both the confinement profile and the field profile can be replaced by more realistic inputs without structural changes to the framework.
The spin-charge coupling is determined by the matrix elements $\langle \psi_i | B_z(x) | \psi_j \rangle$, which sample the full spatial profile of the field weighted by the wavefunction density rather than only the field values at the well minima. 
This is why the linear and step profiles yield different Rabi frequencies even for the same nominal field difference between the wells, particularly when the wavefunctions are delocalized. A key finding of this comparison is that the standard definition of the transverse gradient strength overestimates the Rabi frequency at low barrier heights, where the wavefunction is not yet well localized at the double-well minima. An adjusted definition, based on the actual wavefunction density peak positions, substantially reduces this discrepancy (see Appendix~\ref{app:gradient_comparisions}). This demonstrates that the microscopic spatial structure of the device  meaningfully affects quantitative predictions for the Rabi frequency, beyond what is accessible from a reduced double-dot model.

The modeling approach presented here provides a flexible framework for exploring the parameter space of flopping-mode spin qubits and for connecting spatial device properties to qubit quality. Such results can provide useful guidelines for optimizing FM qubit devices with respect to both single-qubit control and two-qubit interactions, and can complement more detailed electrostatics simulations for specific gate layouts. In future work, the framework could be extended to include additional effects such as charge noise,~\cite{Yoneda2018} more accurate micromagnet field profiles,~\cite{Kawakami2016} valley states in silicon,~\cite{Losert2025} optimized gating schemes,~\cite{Bendersky_2026} and coupling to microwave cavities,~\cite{Mi2018, Borjans2020} enabling a more comprehensive description of flopping-mode qubit devices. In particular, extending the framework to include transverse 
confinement directions would allow for a more complete treatment of valley physics and interface effects in silicon, which are known to influence spin relaxation and spin-charge hybridization in realistic devices. The inclusion of electrostatic disorder and device-to-device variability would further strengthen the connection between the model predictions and experimentally observed qubit performance distributions in large arrays.

\section{Conclusion}
\label{sec:conclusion}
In this work, we applied a flexible framework for modeling flopping-mode spin qubits in double-well confinement potentials with spatially varying magnetic fields. By retaining the spatial structure of the device while remaining computationally efficient, the approach bridges the gap between effective low-energy models and a more detailed finite-element-method simulation of a full gate stack. This allows us to directly relate microscopic device parameters, such as double-well separation and barrier height, to experimentally relevant qubit quality metrics. 

By using this framework, we investigated electrically driven single-qubit dynamics based on electric dipole spin resonance. Our simulations reveal a clear tradeoff between achievable Rabi frequencies and the spectral purity of the corresponding oscillations. While strong charge hybridization enhances the electric dipole moment and enables faster Rabi oscillations, it simultaneously increases spin-charge admixture and leads to leakage into other orbital states, thereby reducing the purity of the Rabi oscillations. Importantly, we find that microscopic device parameters such as the double-well separation can significantly influence this balance, highlighting how device geometry can affect qubit performance.

We further extended the analysis to two-qubit systems and investigated the emergence of capacitive exchange coupling between two neighboring flopping-mode qubits. By using an effective two-qubit Hamiltonian derived from the microscopic model, we evaluated the exchange strength as a function of barrier height, magnetic field gradient strength, and device geometry.
Our analysis reveals the dependence of exchange strength on both intra-qubit and inter-qubit length scales, demonstrating the importance of a microscopic device geometry-sensitive treatment in determining two-qubit coupling strengths.

Additionally, we find that effective double-dot models can fail both quantitatively and qualitatively in certain parameter regimes. The standard step-gradient approximation overestimates the Rabi frequency by up to $\sim$20\% at low barrier heights, and the perturbative double-dot expression for the capacitive exchange coupling develops an unphysical divergence that is absent in the spacially resolved calculation. These results underscore the value of retaining the full spatial structure of the device in qubit modeling.

\section*{Acknowledgments}
 MB acknowledges financial support from the University of Augsburg through seed funding project 2023-26. 

\appendix
\section{Harmonic oscillator basis functions}
\label{app:hermite_polynomials}
In our microscopic modeling approach, the wavefunctions are expanded over one-dimensional harmonic oscillator eigenstates. These functions are expressed in terms of Hermite polynomials and provide a convenient basis for evaluating matrix elements of $H_\mathrm{1FM}$ and constructing the basis functions of $H_\mathrm{2FM}$ analytically.

The normalized harmonic oscillator eigenfunctions are given by:
\begin{equation}
    \phi_n(x) \equiv \frac{1}{(2^n n! \sqrt{\pi} \, \ell)^{1/2}} \,
    H_n\!\left(\frac{x}{\ell}\right)
    \exp\!\left(-\frac{x^2}{2\ell^2}\right),
    \label{HO_basis}
\end{equation}
where $H_n(x)$ denotes the $n$th Hermite polynomial and $\ell \equiv \sqrt{\hbar/(m\omega)}$ is the characteristic length of the harmonic oscillator, associated with an effective confinement frequency $\omega$, which we adjust to the local curvature of the double-well potential minima (see Eq.~\eqref{eq:def-ell}). This gives $\ell \propto (d_\mathrm{LR}^2/\varepsilon_\mathrm{b})^{1/4}$. These functions form an orthonormal basis satisfying
\begin{equation}
    \int_{-\infty}^{\infty} \phi_n(x)\phi_m(x)\,dx = \delta_{nm}.
\label{eq:HO_orthogonality}
\end{equation}
Several identities of the Hermite polynomials are used to evaluate matrix elements efficiently. The Hermite polynomials satisfy the recurrence relation:
\begin{equation}
    H_{n+1}(x) = 2x H_n(x) - 2n H_{n-1}(x),
\label{eq:hermite_recursion}
\end{equation}
and the derivative identity:
\begin{equation}
    \frac{d}{dx} H_n(x) = 2n H_{n-1}(x).
\label{eq:hermite_derivative}
\end{equation}
Using these properties, the action of the position operator on the harmonic oscillator basis can be written as:
\begin{equation}
x\,\phi_n(x)
=
\frac{\ell}{\sqrt{2}}
\left(
\sqrt{n+1}\,\phi_{n+1}(x)
+
\sqrt{n}\,\phi_{n-1}(x)
\right),
\label{eq:x_operator}
\end{equation}
which allows matrix elements to be evaluated algebraically without performing explicit integrals as seen from the example below.

As an illustrative example, we evaluate the matrix element of the position operator between two harmonic oscillator states,
\begin{equation}
    \langle n|x|m\rangle =
    \int_{-\infty}^{\infty} \phi_n(x)\, x \,\phi_m(x)\,dx .
\end{equation}
Using the identity in Eq.~(\ref{eq:x_operator}) acting on $\phi_m(x)$ gives
\begin{equation}
    x\phi_m(x)
    =
    \frac{\ell}{\sqrt{2}}
    \left(
    \sqrt{m+1}\,\phi_{m+1}(x)
    +
    \sqrt{m}\,\phi_{m-1}(x)
    \right).
\end{equation}
Substituting this expression into the matrix element yields
\begin{equation}
    \langle n|x|m\rangle
    =
    \frac{\ell}{\sqrt{2}}
    \left(
    \sqrt{m+1}\,\langle n|m+1\rangle
    +
    \sqrt{m}\,\langle n|m-1\rangle
    \right).
\end{equation}
Using the orthonormality relation in Eq.~(\ref{eq:HO_orthogonality}), we obtain
\begin{equation}
    \langle n|x|m\rangle
    =
    \frac{\ell}{\sqrt{2}}
    \left(
    \sqrt{m+1}\,\delta_{n,m+1}
    +
    \sqrt{m}\,\delta_{n,m-1}
    \right).
\end{equation}
The same algebraic approach extends to all terms appearing in $H_\mathrm{1FM}$. The kinetic energy operator, expressed via the momentum operator $p = i\hbar(\sqrt{n}\,|n-1\rangle\langle n| -\sqrt{n+1}\,|n+1\rangle\langle n|)/(\sqrt{2}\,\ell)$, connects only neighboring basis states. The quartic double-well confinement potential $V(x)$, being a polynomial in $x$, is handled by repeated application of Eq.~\eqref{eq:x_operator}. Likewise, the detuning term (linear in $x$) and the magnetic field gradient (also linear in $x$) each produce matrix elements expressible in closed form via Eq.~\eqref{eq:x_operator}. The result is that $H_\mathrm{1FM}$ is represented as a sparse
$n_\mathrm{max} \times n_\mathrm{max}$ banded matrix, which can be diagonalized efficiently to yield the single-particle spectrum and spatially resolved wavefunctions. In our simulations we use $n_\mathrm{max} = 50$, at which point the FM qubit energies and wavefunctions are well converged. A quantitative demonstration of the convergence of the lowest eigenenergies, EDSR drive matrix element, and exchange strength with $n_\mathrm{max}$ is provided in Appendix~\ref{app:convergence}.

\section{Convergence checks}
\label{app:convergence}
This Appendix presents three convergence and validation 
checks for the numerical framework: (i) convergence of the FM qubit 
eigenenergies, EDSR drive matrix element, and exchange strength with 
harmonic-oscillator basis size $n_\mathrm{max}$; (ii) sensitivity of the 
exchange coupling to the Coulomb regularization parameter $\alpha$; and 
(iii) a physical argument justifying the truncation of the CI basis to 
the two lowest orbital states per qubit.

\begin{figure}[htb]
    \centering
    \includegraphics[width=\columnwidth]{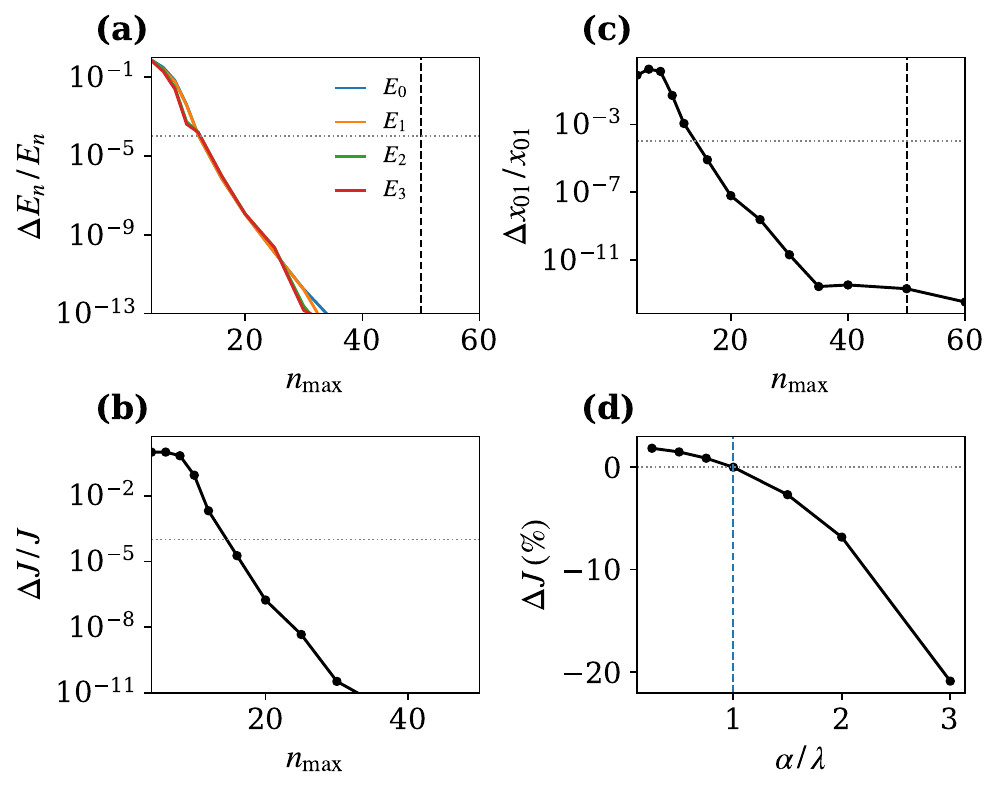}
    \caption{\textbf{Convergence and sensitivity checks}.
    \textbf{(a)-(c)} Relative error of \textbf{(a)} the four lowest FM qubit eigenenergies $E_n$, \textbf{(b)} the exchange strength $J$ and \textbf{(c)} the EDSR drive matrix element $x_{01} = \langle\psi_1|x|\psi_0\rangle$ as a function of the harmonic oscillator basis size.
    \textbf{(d)} Sensitivity of the exchange strength $J$ with respect to the Coulomb regularization parameter $\alpha$, with basis size $n_\text{max} = 50$ fixed. The blue dashed line marks the physical choice $\alpha = \lambda$ used in all calculations, where $\lambda$ is the harmonic-oscillator length of the dot.}
    \label{fig:convergence}
\end{figure}

Figure~\ref{fig:convergence} shows the relative error in the four lowest FM-qubit eigenenergies ($E_0$-$E_3$), the EDSR drive matrix element, which is proportional to the Rabi frequency $f_\mathrm{R}$, and the exchange strength $J$ as a function of basis size $n_\mathrm{max}$, relative to well-converged reference calculations ($n_\mathrm{ref} = 100$ for panels~(a),(c); $n_\mathrm{ref} = 80$ for panel~(b) with reference value $J(n_\mathrm{ref}) = 5.71~\mu$eV). 
All three quantities cross the $10^{-4}$ relative-error threshold by $n_\mathrm{max} \approx 12 \-- 16$ and reach errors below $10^{-10}$ at $n_\mathrm{max} = 50$.
The choice used throughout the paper is therefore conservative with respect to any numerical errors. We also show the sensitivity of the calculated exchange coupling to the Coulomb regularization parameter $\alpha$ in panel (d). $J$ varies by less than 3\% when $\alpha$ is changed by a factor of 1.5 in either direction, confirming that the results are insensitive to the exact value of $\alpha$.

The truncation to the two lowest orbital states ($n = 0, 1$) per qubit used in the restricted CI bases for modeling two flopping mode qubits is justified by two arguments. 
First, the energy gap from the top of the computational subspace $\{n=0,1\}\otimes\{\uparrow,\downarrow\}$ to the first excluded state ($n=2$), which we refer to as the isolation gap, ranges from $\sim 400 \,\text{\textmu eV}$ at $\varepsilon_\mathrm{b} = 0.5\,\text{meV}$ to $\sim 1100\,\text{\textmu eV}$ at $\varepsilon_\mathrm{b} = 2\,\text{meV}$ (see Fig.~\ref{fig:DWP_orbital_structure}(a)). 
This exceeds the capacitive coupling $\kappa \approx 20 \-- 40 \,\text{\textmu eV}$ by a factor of $10 \-- 55$ and the spin splitting $\Delta E_\mathrm{spin} = 23.2\,\text{\textmu eV}$ by a similar margin throughout the parameter range studied. 
Virtual transitions from the computational subspace to $n = 2$ are therefore strongly suppressed perturbatively.
Second, in the capacitive regime ($d_\mathrm{FM} \gg \lambda$), writing the electron coordinates as $x_{\mathrm{a},\mathrm{b}} = \mp d_\mathrm{FM}/2 + \delta x_{\mathrm{a},\mathrm{b}}$, the Coulomb interaction can be expanded in powers of $\lambda/d_\mathrm{FM}$. 
\begin{equation}
    V_\mathrm{C}(x_\mathrm{a}, x_\mathrm{b}) \approx \frac{e^2}{4\pi\epsilon\,d_\mathrm{FM}}
    \left[1 + \frac{\delta x_\mathrm{a} - \delta x_\mathrm{b}}{d_\mathrm{FM}} 
    + \left(\frac{\delta x_\mathrm{a} - \delta x_\mathrm{b}}{d_\mathrm{FM}}\right)^2 
    + \cdots \right],
\label{eq:multipole}
\end{equation}
The dipole term ($\propto \delta x_{\mathrm{a},\mathrm{b}}$) governs the intra-subspace coupling $\kappa$ via the selection rule $\langle n'|\delta x|n\rangle \propto \lambda(\sqrt{n+1}\,\delta_{n',n+1} + \sqrt{n}\,\delta_{n',n-1})$, but also couples $n=1$ to $n=2$ outside the subspace at order $\lambda/d_\mathrm{FM} \lesssim 0.09$.
A numerical evaluation of the Coulomb matrix elements at $\varepsilon_\mathrm{b} = 1$~meV confirms $|M_2|/|M_1| = 12\%$, consistent with this scaling, and yields a second-order correction $\delta J \approx |M_2|^2/\Delta_\mathrm{orb}(n=2) \approx 0.19~\mu$eV, less than 1\% of $\kappa$ throughout the parameter range studied.

\section{Double-dot model for a flopping-mode qubit}
\label{app:low_energy_models}
The double-dot model, introduced in Refs.~\onlinecite{Benito2019, Croot2020} to study a single FM qubit, retains only the two lowest orbital states of a double-well potential instead of a full spatial description as in Eq.~\eqref{H_FM}. Those two states are the symmetric (bonding) $|+\rangle$ and antisymmetric (antibonding) $|-\rangle$ orbital states ($\ket{\pm} \equiv (\ket{\mathrm{R}} \pm \ket{\mathrm{L}})/\sqrt{2}$ with $\ket{\mathrm{L}}$ and $\ket{\mathrm{R}}$ representing the lowest-energy orbital state of the left and right dot, respectively), together with spin-1/2. The double-dot model Hamiltonian $H_\mathrm{DD}$ can be written as
\begin{equation}
    H_\mathrm{DD} =  t_\mathrm{c}\tau_x - \varepsilon_\mathrm{d} \tau_z + \frac{g \mu_\mathrm{B} B_x}{2}\sigma_x - \frac{g\mu_\mathrm{B} \Delta B_z}{2}\sigma_z\tau_z,
    \label{eq:H2s}
\end{equation}
where $\tau_x \equiv |\mathrm{L}\rangle\langle \mathrm{R}| + |\mathrm{R} \rangle\langle \mathrm{L}|$ and $\tau_z \equiv |\mathrm{L} \rangle\langle \mathrm{L}| - |\mathrm{R}\rangle\langle \mathrm{R}|$ are Pauli matrices in the left-right orbital subspace, $t_\mathrm{c}$ is the interdot tunnel coupling, and $\Delta B_z$ denotes the transverse gradient strength. The bonding-antibonding splitting $2|t_\mathrm{c}|$ sets the orbital energy scale, and the spin Rabi frequency to leading order in $\Delta B_z$ is given by $\Omega_\mathrm{s} \approx 2 t_\mathrm{c} g \mu_\mathrm{B} \Delta B_z \Omega_\mathrm{c}/|4 t_\mathrm{c}^2 - E_x^2|$,~\cite{Benito2019} where $\Omega_\mathrm{c} = (4 t_\mathrm{c}^2 + \varepsilon_\mathrm{d}^2)^{1/2}$ and $E_x \equiv g \mu_\mathrm{B} B_x$. This is a compact and analytically tractable description that captures the essential physics of spin-charge hybridization. However, the double-dot model treats $t_\mathrm{c}$ as a free input parameter. Our microscopic model provides a direct link from a specific device geometry (e.g. confinement potential) to the four-dimensional low-energy subspace of the FM qubit and the relevant model parameters (such as tunnel coupling, see Fig.~\ref{fig:DWP_orbital_structure}(b)), while retaining the spatial properties of the low-energy states (see Fig.~\ref{fig:FM_spatialprops}, for example).

In Ref.~\onlinecite{Cayao2020}, the authors study two capacitively-coupled FM qubits, including for a lined-up configuration as shown in Fig.~\ref{fig:FM_qubits}(a), built up from two double-dot models. The Coulomb interaction between electrons on different double dots is parameterized through three configuration-dependent repulsion energies: $U_\mathrm{N}$, describing the case where both electrons occupy the inner (near) dots; $U_\mathrm{F}$, for both in the outer (far) dots; and $U_\mathrm{M}$, for the intermediate case where both electrons sit on the same side (left or right) in their double dot. These three values enter the two-qubit Hamiltonian both as a single-particle detuning shift $\propto (U_\mathrm{N} - U_\mathrm{F})$ and as an orbital-orbital coupling $\kappa = (2 U_\mathrm{M} - U_\mathrm{F} - U_\mathrm{N})/4$ that mediates the effective qubit-qubit interaction:
\begin{equation}
    H_\mathrm{C} = -\frac{U_\mathrm{N} - U_\mathrm{F}}{4} \tau_z^{(\mathrm{A})} + \frac{U_\mathrm{N} - U_\mathrm{F}}{4} \tau_z^{(\mathrm{B})} + \kappa \tau_z^{(\mathrm{A})} \tau_z^{(\mathrm{B})},
\end{equation}
with $(\mathrm{A})$ and $(\mathrm{B})$ denoting the operators acting on the double-dot basis states of FM qubit A and B, respectively.

\begin{figure}[htb]
    \includegraphics[width=\linewidth]{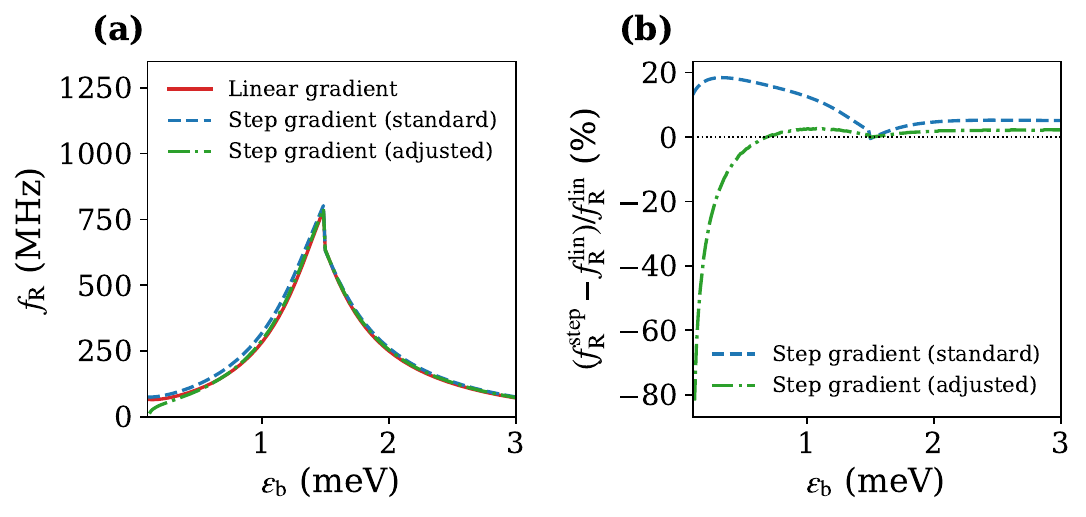}
    \caption{
        \textbf{Comparison of Rabi frequencies}.
        \textbf{(a)} The Rabi frequency as a function of barrier height $\varepsilon_\mathrm{b}$, as obtained with the microscopic model with linear magnetic field gradient profile and the double-dot model, considering the standard ($\Delta B_x = b_z d_\mathrm{LR}/2$) and adjusted (see Eq.~\eqref{eq:Bstep_eff}) transverse gradient strength.
        \textbf{(b)} The relative difference between the Rabi frequencies obtained with the microscopic model and those obtained with the double-dot model, with standard and adjusted transverse gradient strength.
    }
    \label{fig:AppC}
\end{figure}

To isolate the exchange interaction from the direct electrostatic energy shift, we can consider the symmetric detuning point $\varepsilon_\mathrm{d} = -(U_\mathrm{N} - U_\mathrm{F})/2$, which compensates the net charge imbalance induced by the Coulomb interaction; the same compensation regime is considered for the microscopic model in Sec.~\ref{sec:model}. In our microscopic framework, however, neither $\kappa$ nor the $U$ parameters appear as inputs: the full distance and wavefunction-dependent Coulomb interaction $V_\mathrm{C}(x_1, x_2) \propto |(x_1 - x_2)^2 + \alpha^2|^{-1/2}$ is evaluated directly from the Hermite polynomial basis, naturally encoding all three charge configurations and their geometry dependence. The coupling $\kappa$ therefore emerges as an effective quantity characterizing the overall scale of the interaction, rather than a model parameter, and the values $\kappa$ extracted in Fig.~\ref{fig:FM_exchange} reflect the geometry and wavefunction overlap of our specific device configurations.

\section{Effective transverse gradient strength}
\label{app:gradient_comparisions}
In a reduced double-dot description of an FM qubit, a continuous magnetic field gradient profile $B_z(x)$ must be replaced by an effective step function $\pm \Delta B_x$ acting on each dot site, with $\Delta B_x$ the transverse gradient strength. For our microscopic model with linear gradient (Eq.~\eqref{b_grad}) and double-well potential with minima at $x=\pm d_\mathrm{LR}/2$ (Eq.~\eqref{eq:potential}), $\Delta B_x = b_z d_\mathrm{LR}/2$ is a natural choice for the transverse gradient strength, evaluating the field at the nominal dot centers $\pm d_\mathrm{LR}/2$. However, this \textit{standard} definition overestimates the effective spin-orbit coupling whenever the electron wavefunction is not sharply localized at $\pm d_\mathrm{LR}/2$ (see Fig.~\ref{fig:FM_qubits}(c)). Based on the microscopic model and spatial properties of the low-energy states, we can define an adjusted transverse gradient strength as follows:
\begin{equation}
     \Delta B_x = b_z (x_\mathrm{R}^\mathrm{max} - x_\mathrm{L}^\mathrm{max})/2,
\label{eq:Bstep_eff}
\end{equation}
with $x_\mathrm{L,R}^\mathrm{max}$ the positions of the wavefunction density peaks in the two wells, as shown in Fig.~\ref{fig:FM_spatialprops}(c). The adjusted transverse gradient strength can deviate significantly from the standard definition, especially when the barrier is low and the two density peaks are not strongly confined in the two wells.

In Fig.~\ref{fig:AppC}, we compare the Rabi frequency as a function of barrier height of three approaches for an EDSR-driven FM qubit at zero detuning: the microscopic model with linear gradient, and the double-dot model with standard and adjusted definition of the transverse (step) gradient strength. We find that all three curves agree well near the spin-to-orbital crossover where the Rabi frequency peaks, but the standard transverse strength overestimates the linear-gradient result by up to ${\sim}20$ percent at low $\varepsilon_\mathrm{b}$. This reflects the discrepancy between $d_\mathrm{LR}/2$ and $(x_\mathrm{R}^\mathrm{max} - x_\mathrm{L}^\mathrm{max})/2$. The adjusted definition substantially reduces this deviation, remaining within ${\lesssim}5$ percent across most of the barrier height range. At very low $\varepsilon_\mathrm{b}$ the effective step itself underestimates the linear-gradient result: here the wavefunction has not yet split into two resolved peaks, so $x_\mathrm{L,R}^\mathrm{max} \ll d_\mathrm{LR}/2$ and the step approximation itself breaks down. These results show that the accuracy of a low-energy double-dot model can be substantially improved by taking into account the spatial properties of the device and its low-energy spectrum, particularly in the intermediate barrier-height regime relevant to qubit operation. Since an arbitrary field profile $B_z(x)$ enters the model only through its matrix elements in the Hermite polynomial basis, which are evaluated exactly, the framework applies to any polynomial field profile without approximation.

\section{Spectral analysis of Rabi oscillations}
\label{app:additional-spectral-properties}
\begin{figure}[tb]
    \includegraphics[width=\linewidth]{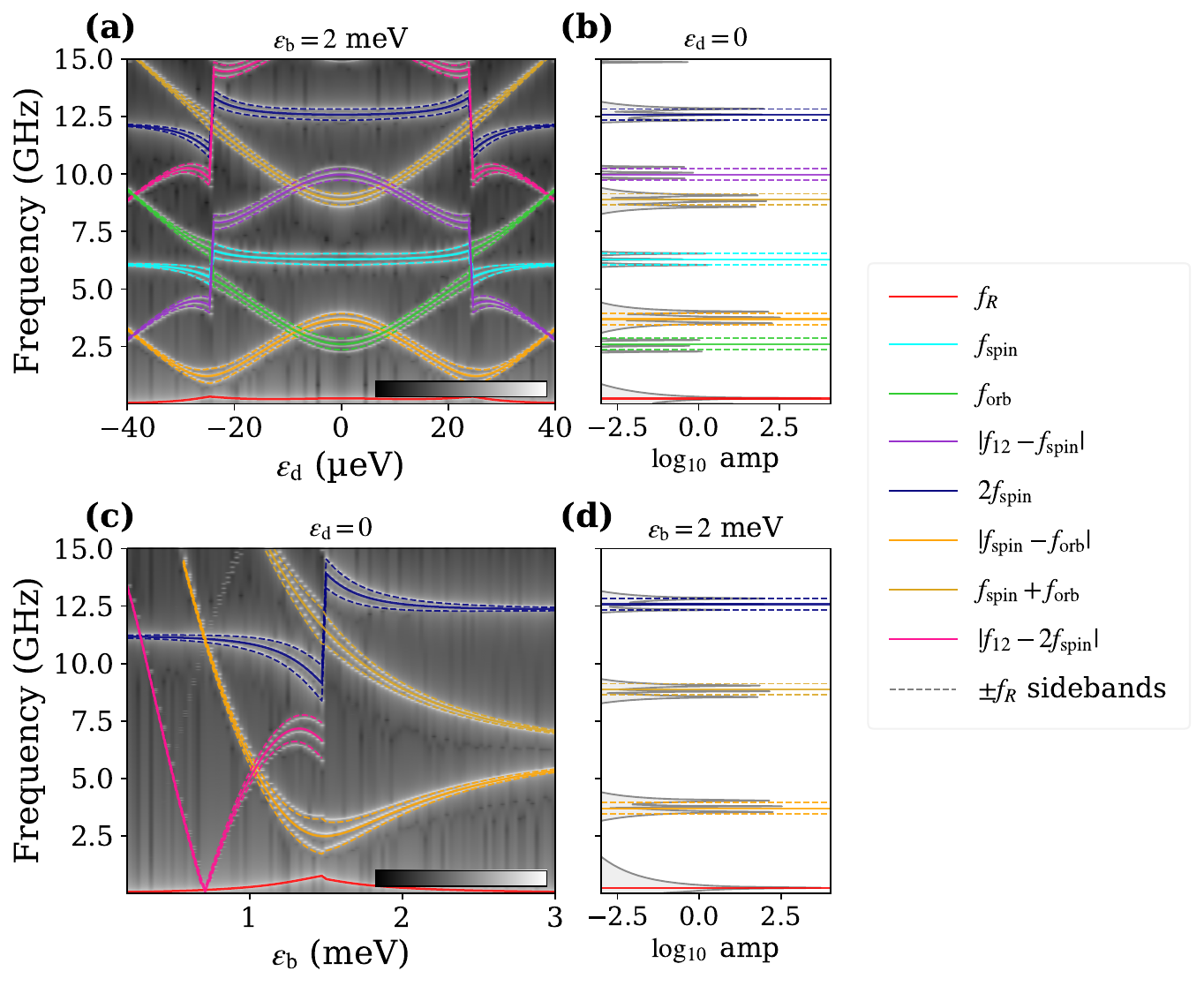}
    \caption{
        \textbf{Spectral analysis of Rabi oscillations}.    
        \textbf{(a)}-\textbf{(d)} The Fourier spectrum of the Rabi oscillations of an EDSR-driven flopping-mode qubit \textbf{(a)} as a function of $\varepsilon_\mathrm{d}$ at $\varepsilon_\mathrm{b} = 2\,\textnormal{meV}$ with \textbf{(b)} linecut for $\varepsilon_\mathrm{d}=0$ and \textbf{(c)} as a function of $\varepsilon_\mathrm{b}$ at $\varepsilon_\mathrm{d}=0$ with \textbf{(d)} linecut for $\varepsilon_\mathrm{b}=2\,\textnormal{meV}$.
    }
    \label{fig:AppD}
\end{figure}

To characterize the EDSR-driven dynamics of the FM qubit beyond the Rabi frequency and spectral purity maps as presented in Fig.~\ref{fig:FM_EDSR_1}, we compute the Fourier transform of the time-domain ground state probability $P_\mathrm{0}(t)$ (see Figs.~\ref{fig:FM_EDSR_1}(a)-(c)). Figure~\ref{fig:AppD} presents two-dimensional FFT amplitude maps as a function of frequency for two parameter sweeps, with color-coded lines marking the key frequencies in the spectrum.

The dominant spectral feature throughout is the Rabi frequency $f_\mathrm{R}$ at the low-frequency end of the spectrum (red line). This arises directly from the sinusoidal envelope of the Rabi oscillation $P(t) \propto \cos^2(\pi f_\mathrm{R} t)$ and is by far the strongest contribution to the FFT amplitude. At higher frequencies, weaker but clearly resolved features appear at the spin transition frequency $f_\mathrm{spin}$ (which is also set as the drive frequency) and its harmonics $2f_\mathrm{spin}$ (blue lines), and $f_\mathrm{orb}$ (green line). These originate from multi-level leakage and corrections beyond the rotating-wave approximation in the multi-orbital single-FM-qubit Hamiltonian. Note that some of these frequencies are absent at zero detuning, making it a symmetric sweet-spot for cleaner Rabi oscillations. 

More physically significant are the sideband pairs $f_\mathrm{spin} \pm f_\mathrm{orb}$ (orange lines). These arise from the spin-orbit coupling induced by the slanting magnetic field gradient, which hybridizes the spin and orbital degrees of freedom and generates combination tones at the sum and difference of the two characteristic frequencies. As shown in panels (b) and (d), these orbital sidebands are the dominant high-frequency contribution after the Rabi harmonics, appearing roughly one to two decades below the Rabi peak.

Another striking parameter dependence is observed in the form of dressed-state sidebands near the spin-to-charge crossover at $\varepsilon_\mathrm{d} \approx \pm 23$~$\mu$eV in panel (a) and $\varepsilon_\mathrm{b} \approx 1.5$~meV in panel (c), where $f_\mathrm{orb} \approx f_\mathrm{spin}$. The Rabi frequency $f_\mathrm{R}$ itself peaks at this spin-to-charge crossover, which maximizes the Autler-Townes splitting of the driven spin. The dotted lines in Fig.~\ref{fig:AppD} show the theoretical positions of the dressed-state sidebands, whose separation from the bare frequency harmonics is widest at the transition. With increasing $\varepsilon_\mathrm{d}$ or $\varepsilon_\mathrm{b}$, the two sidebands converge symmetrically towards the single frequency from which they split. 

The spectral content of $P_0(t)$ also motivates a frequency-domain formulation of the spectral purity introduced in Sec.~\ref{sec:EDSR}. 
Defining $\mathcal{P}_\mathrm{FFT}$ as the fraction of total Fourier spectral power contained in the dominant Rabi peak,
\begin{equation}
\mathcal{P}_\mathrm{FFT} \equiv 
\frac{\displaystyle\sum_{|f - f_R| < \delta f} |\tilde{A}(f)|^2}
{\displaystyle\sum_{f > 0} |\tilde{A}(f)|^2},
\label{eq:PFFT_appD}
\end{equation}
where $\tilde{A}(f)$ is the Fourier amplitude of the mean-subtracted $P_0(t)$ and $\delta f$ is a window half-width enclosing the Rabi peak while excluding leakage features, this measure directly quantifies the fraction of oscillation energy that follows the ideal single-frequency Rabi pattern. 
The FFT maps of Fig.~\ref{fig:AppD} make clear why $\delta f$ can be chosen freely within a wide range.
The Rabi peak ($f_R \lesssim 0.8$~GHz) is separated from the nearest leakage feature ($f_\mathrm{spin} \gtrsim 2$~GHz) by a spectral gap of at least 1.2~GHz, so the result is entirely independent of the precise window width. 
Numerically, we verified this by computing $\mathcal{P}_\mathrm{FFT}$ for $\delta f \in \{0.15, 0.30, 0.50\}$~GHz on two linecuts through the parameter space, obtaining identical results in all cases.

\begin{figure}[tb]
    \centering
    \includegraphics[width=\linewidth]{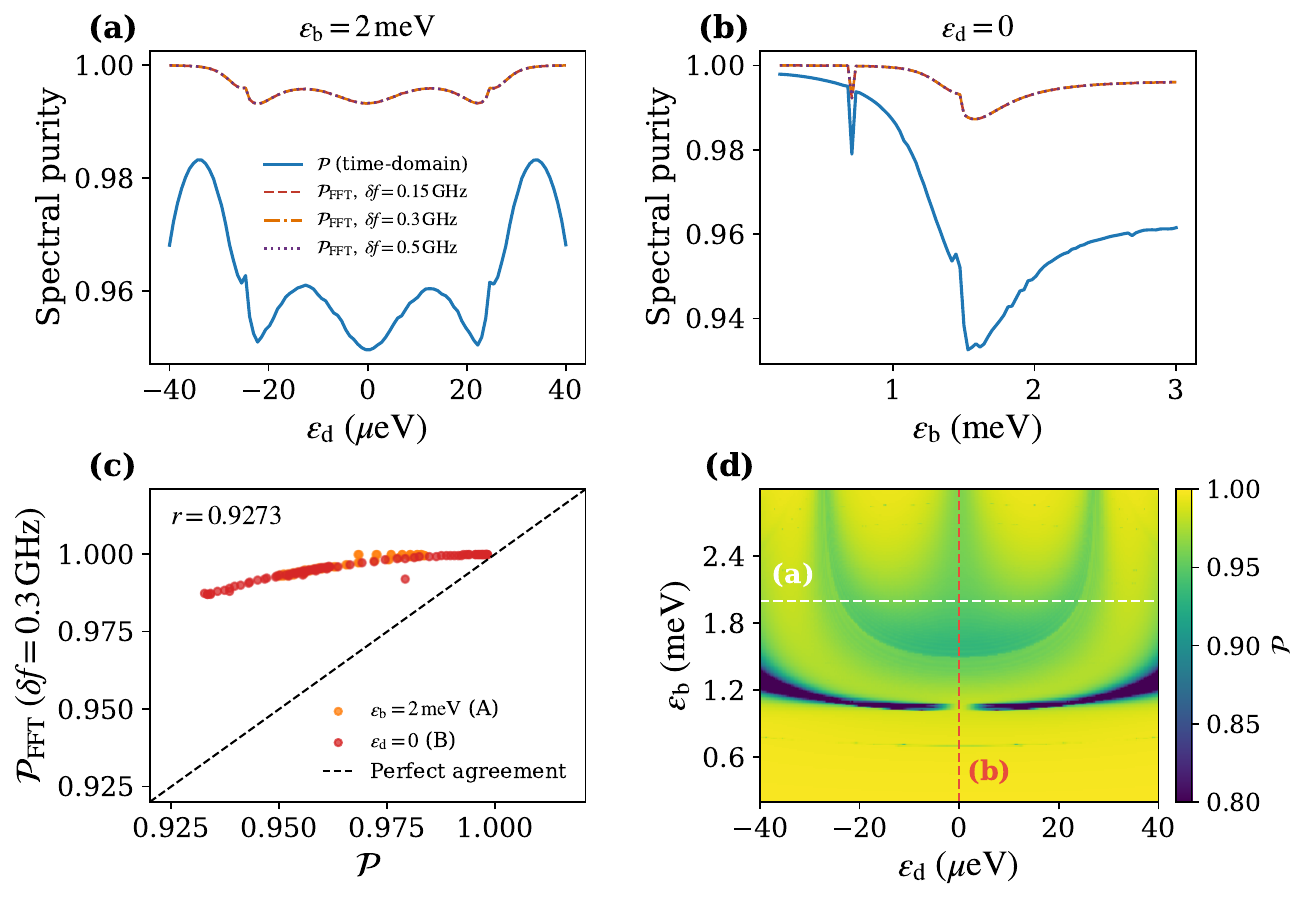}
    \caption{\textbf{Comparison of time-domain ($\mathcal{P}$) and frequency-domain ($\mathcal{P}_\mathrm{FFT}$) spectral purity measures}.
    \textbf{(a),(b)} Spectral purity as a function of \textbf{(a)} $\varepsilon_\mathrm{d}$ with $\varepsilon_\mathrm{b} = 2$~meV (linecut~A), and of (\textbf{(b)} $\varepsilon_\mathrm{b}$ with $\varepsilon_\mathrm{d} = 0$ (linecut~B).
    Three coloured curves correspond to $\delta f \in \{0.15, 0.30, 0.50\}$~GHz and are indistinguishable, confirming window-width independence.
    \textbf{(c)}~Scatter plot of $\mathcal{P}$ versus $\mathcal{P}_\mathrm{FFT}$ of all 200 combined data points of linecuts A and B against the perfect-agreement diagonal; Pearson $r$ indicated.
    \textbf{(d)}~Full time-domain $\mathcal{P}$ map ($d_\mathrm{LR} = 100$~nm), with linecuts A and B indicated.}
    \label{fig:purity_comparison}
\end{figure}

The time-domain purity $\mathcal{P}$ (Eq.~(\ref{eq:spectral-purity})) and $\mathcal{P}_\mathrm{FFT}$ are equivalent in physical content. 
By Parseval's theorem, the mean-square deviation of $P_0(t)$ from the ideal Rabi profile equals the total leakage spectral power, so both measures encode the same ratio of leakage power to Rabi power. For a leakage signal dominated by a few sinusoidal components, as confirmed by the FFT maps in Fig.~\ref{fig:AppD}, the $L^1$ and $L^2$ norms are related by $\langle|x|\rangle \approx (2/\pi)\langle x^2\rangle^{1/2}$, making the two measures proportional in the high-purity limit. 
A numerical comparison on two linecuts through the parameter space 
(linecut~A: $\varepsilon_\mathrm{b} = 2$~meV varying $\varepsilon_\mathrm{d}$; linecut~B: $\varepsilon_\mathrm{d} = 0$ varying $\varepsilon_\mathrm{b}$; 200 data points total) confirms a Pearson correlation of $r \approx 0.93$--$0.97$ between the two definitions (see Fig.~\ref{fig:purity_comparison}). 
The systematic offset ($\langle\mathcal{P}\rangle \approx 0.965$ vs.\ $\langle\mathcal{P}_\mathrm{FFT}\rangle \approx 0.996$) is fully accounted for by the $L^1$/$L^2$ norm ratio: the FFT measure reports that only $\sim$0.4\% of the spectral power lies outside the Rabi peak, while the time-domain measure sees a $\sim$3.5\% deviation in the mean oscillation shape, consistent with the expected factor of $2/\pi \approx 0.64$. 
Both measures identify the same regions of high and low spectral 
purity and lead to identical physical conclusions.

\section{Notation table}
\label{app:table}
Table.~\ref{tab:notation} provides a list of all the model parameters used along with their units, default values if any in parenthesis and their physical meanings.

\begin{table}[htb]
\centering
\footnotesize
\setlength{\tabcolsep}{4pt}
\caption{Model parameters used throughout the paper.}
\label{tab:notation}
\begin{tabular}{lll}
\toprule
Symbol & Units & Meaning \\
\midrule
\multicolumn{3}{l}{\textit{Single-qubit}} \\
\midrule
$m_\mathrm{e}^\ast$ & $m_\mathrm{e}$ & Effective electron mass \\
$g$ & --- (2)& Land\'{e} g-factor \\
$B_x$ & T (0.2) & Homogeneous field \\
$b_z$ & mT/nm (2) & Gradient strength \\
$B_z(x)$ & T & Field gradient profile \\
$d_\mathrm{LR}$ & nm (100)& Well separation \\
$\varepsilon_\mathrm{b}$ & meV & Barrier height \\
$\varepsilon_\mathrm{d}$ & \textmu eV & Inter-well detuning \\
$\lambda$ & nm & Dot size, Eq.~\ref{eq:def-ell} \\
$\ell$ & nm & Hermite basis length \\
$t_\mathrm{c}$ & \textmu eV & Tunnel coupling \\
$\Delta E_\mathrm{orb}$ & \textmu eV & Orbital splitting \\
$\Delta E_\mathrm{spin}$ & \textmu eV & Spin splitting \\
$f_\mathrm{R}$ & MHz & Rabi frequency \\
$\mathcal{P}$ & --- & Spectral purity, Eq.~\ref{eq:spectral-purity} \\
\midrule
\multicolumn{3}{l}{\textit{Two-qubit}} \\
\midrule
$d_\mathrm{FM}$ & nm (250)& Inter-qubit distance \\
$\kappa$ & \textmu eV & Capacitive coupling, Eq.~\ref{eq:J_analytic} \\
$J$ & \textmu eV & Exchange strength \\
$J_\mathrm{DD}$ & \textmu eV & Exchange, double-dot model \\
$E_z$ & \textmu eV & Gradient Zeeman splitting \\
$E_x$ & \textmu eV & Homogeneous Zeeman splitting \\
$\alpha$ & nm & Coulomb cutoff \\
$U_\mathrm{N,F,M}$ & \textmu eV & Coulomb repulsion energies \\
\midrule
\multicolumn{3}{l}{\textit{Basis and numerical}} \\
\midrule
$n_\mathrm{max}$ & --- (50) & Basis size \\
$\phi_n(x)$ & nm$^{-1/2}$ & HO eigenfunction \\
$H_n(x)$ & --- & Hermite polynomial \\
\bottomrule
\end{tabular}
\end{table}
\FloatBarrier

\bibliography{references}

\end{document}